\documentclass[10pt,journal,letterpaper,twosides,twocolumns]{IEEEtran}

\usepackage{etex}

\usepackage{amsmath,amssymb,amscd,latexsym,dsfont,wasysym,bbm}
\usepackage{float}
\usepackage{color}
\usepackage{graphicx,subfigure}
\usepackage{multirow,multicol} 
\usepackage{verbatim}
\usepackage{psfrag}
\usepackage{comment}
\usepackage{makeidx}
\usepackage[]{algorithm}
\usepackage{algorithmic}
\usepackage{cite}
\usepackage{cases}
\usepackage{amsthm}
 \usepackage{setspace}
\usepackage[table]{xcolor}
 \usepackage{pgfplots} 
\newcommand{\tr}[1]{\textrm{#1}}
\newcommand{\mr}[1]{\mathrm{#1}}
\newcommand{\tnr}[1]{{\textnormal{#1}}}

\newcommand{\mc}[1]{\mathcal{#1}}
\newcommand{\mf}[1]{\mathsf{#1}}
 
\newcommand{\ms}[1]{\mathds{#1}}

\newcommand{\ov}[1]{\overline{#1}}

\newcommand{\bb}{\boldsymbol{b}}

\newcommand{\bx}{\boldsymbol{x}}

\newcommand{\by}{\boldsymbol{y}}

\newcommand{\bz}{\boldsymbol{z}}

\newcommand{\seqref}[1]{Eq.~(\ref{#1})}  
\newcommand{\figref}[1]{Fig.~\ref{#1}}

\newcommand{\secref}[1]{Sec.~\ref{#1}}

\newcommand{\exref}[1]{Example~\ref{#1}}

\newcommand{\ie}{i.e.,~} 		
\newcommand{\eg}{e.g.,~}	
\newcommand{\cf}{cf.~}		

\newcommand{\argmax}{\mathop{\mr{argmax}}}

\newcommand{\set}[1]{\{#1\}}
\newcommand{\SET}[1]{\left\{#1\right\}}

\newcommand{\cd}{\cdot}
\newcommand{\ld}{\ldots}

\newcommand{\PR}[1]{\Pr\SET{#1}}       	
\newcommand{\pdf}{\mf{p}}            			

\newcommand{\IND}[1]{\ms{I}\big[{#1}\big]}   	
\newcommand{\Ex}{\ms{E}}     			

\newcommand{\mcF}{\mc{F}}

\newcommand{\mcR}{\mc{R}}

\newcommand{\mcX}{\mc{X}}

\newcommand{\mfa}{\mf{a}}

\newcommand{\mfm}{\mf{m}}

\newcommand{\mfs}{\mf{s}}

\newcommand{\mfI}{\mf{I}}
\newcommand{\mfJ}{\mf{J}}

\newcommand{\mfR}{\mf{R}}

\newcommand{\Real}{\mathbb{R}}		
\newcommand{\Binary}{\mathbb{B}}	

\newcommand{\CSI}{\mathsf{csi}}  
\newcommand{\corSNR}{\delta}  
\newcommand{\fD}{f_\text{D}}  

\newcommand{\SNR}{\mathsf{snr}}  
\newcommand{\SNRrv}{\mathsf{SNR}}  
\newcommand{\SNRav}{\ov{\mathsf{snr}}}  

\newcommand{\PER}{\mr{PER}}  

\newcommand{\X}{\mcX}	

\newcommand{\R}{R}           	
\newcommand{\Rc}{R_\tnr{c}}  	
\newcommand{\Ns}{{\mathop{N_\tnr{s}}}} 	

\newcommand{\IR}{\tnr{ir}}
\newcommand{\VL}{\tnr{VL}}  

\newcommand{\amc}{\tnr{amc}}

\newcommand{\Err}{\mf{ERR}}
\newcommand{\Errb}[1]{\mf{ERR}^{\tnr{b}}_{#1}}       
\newcommand{\kmax}{K}		
\newcommand{\kk}{k}
\newcommand{\Lc}{\tr{L}} 
\newcommand{\xp}{\tnr{xp}}

\newcommand{\Nsk}[1]{\mathop{N_{\tnr{s},#1}}} 	
\newcommand{\Rs}[1]{\mathop{\rho_{#1}}}       

\newcommand{\dBval}{\tnr{dB}} 
\newcommand{\bpcu}{\tnr{[bit/cu]}} 
\newcommand{\bpcuval}{\tnr{bit/cu}} 

\usepackage[acronym,nonumberlist]{glossaries} 
\usepackage[draft]{hyperref}
\newacronym[\glsshortpluralkey=PDFs,\glslongpluralkey=probability density functions]{pdf}{PDF}{probability density function}
\newacronym[\glsshortpluralkey=CDFs,\glslongpluralkey=cumulative density functions]{cdf}{CDF}{cumulative density function}
\newacronym[\glsshortpluralkey=CCDFs,\glslongpluralkey=complementary cumulative density functions]{ccdf}{CDF}{complementary cumulative density function}
\newacronym[\glsshortpluralkey=PMFs,\glslongpluralkey=probability mass functions]{pmf}{PMF}{probability mass function}
\newacronym[]{lhs}{l.h.s.}{left-hand side}
\newacronym[]{rhs}{r.h.s.}{right-hand side} 

\newacronym[]{bicm}{BICM}{bit-interleaved coded modulation}
\newacronym[]{bicmid}{BICM-ID}{BICM with iterative demapping}
\newacronym[]{cm}{CM}{coded modulation}
\newacronym[]{tcm}{TCM}{trellis-coded modulation}
\newacronym[]{mlc}{MLC}{multi-level coding}
\newacronym[]{pam}{PAM}{pulse amplitude modulation}
\newacronym[]{bpsk}{BPSK}{binary phase shift keying}
\newacronym[]{qam}{QAM}{quadrature amplitude modulation}
\newacronym[]{16qam}{16-QAM}{16-points quadrature amplitude modulation}
\newacronym[]{psk}{PSK}{phase shift keying}
\newacronym[\glsshortpluralkey=LLRs,\glslongpluralkey=logarithmic likelihood ratios]{llr}{LLR}{logarithmic likelihood ratio}
\newacronym[]{map}{MAP}{maximum a posteriori}
\newacronym[]{ml}{ML}{maximum likelihood}
\newacronym[]{ep}{EP}{expectation propagation}
\newacronym[\glsshortpluralkey=MIs,\glslongpluralkey=mutual informations]{mi}{MI}{mutual information}
\newacronym[\glsshortpluralkey=GMIs,\glslongpluralkey=generalized mutual informations]{gmi}{GMI}{generalized mutual information}
\newacronym[]{eesm}{EESM}{exponential effective-SNR-mapping}
\newacronym[]{bicm-gmi}{BICM-GMI}{BICM generalized mutual information}
\newacronym[]{awgn}{AWGN}{additive white Gaussian noise}
\newacronym[]{bsc}{BSC}{binary symetric channel}
\newacronym[]{amc}{AMC}{adaptive modulation and coding}
\newacronym[]{csi}{CSI}{channel state information}
\newacronym[]{cqi}{CQI}{channel quality indicator}

\newacronym[]{sp}{SP}{set-partitioning}
\newacronym[]{gsm}{GSM}{global system for mobile communications}
\newacronym[]{edge}{EDGE}{enhanced data rates for GSM evolution}
\newacronym[]{3gpp}{3GPP}{3rd generation partnership project}
\newacronym[]{umts}{UMTS}{Universal Mobile Telecommunication System}
\newacronym[]{lte}{LTE}{Long Term Evolution}
\newacronym[]{dvb}{DVB}{digital video broadcasting}
\newacronym[]{fdd}{FDD}{Frequency Division Duplexing}

\newacronym[\glsshortpluralkey=CCs,\glslongpluralkey=convolutional codes]{cc}{CC}{convolutional code}
\newacronym[\glsshortpluralkey=PCCCs,\glslongpluralkey=parallel concatenated convolutional codes]{pccc}{PCCC}{parallel concatenated convolutional code}
\newacronym[\glsshortpluralkey=TCs,\glslongpluralkey=turbo codes]{tc}{TC}{turbo code}
\newacronym{ldpc}{LDPC}{low-density parity-check}
\newacronym[]{ofdm}{OFDM}{orthogonal frequency-division multiplexing}
\newacronym[]{bep}{BEP}{bit-error probability}
\newacronym[]{wep}{WEP}{word-error probability}
\newacronym[]{sep}{SEP}{symbol-error probability}
\newacronym[]{pep}{PEP}{pairwise-error probability}
\newacronym[]{ttcm}{TTCM}{turbo-trellis coded modulation}
\newacronym[]{uep}{UEP}{unequal error protection}
\newacronym[\glsshortpluralkey=CENCs,\glslongpluralkey=convolutional encoders]{cenc}{CENC}{convolutional encoder}
\newacronym[]{mimo}{MIMO}{multiple-input multiple-output}
\newacronym[\glsshortpluralkey=SNRs,\glslongpluralkey=signal-to-noise ratios]{snr}{SNR}{signal-to-noise ratio}
\newacronym[\glsshortpluralkey=SINRs,\glslongpluralkey=the signal-to-interference-plus-noise ratios]{sinr}{SINR}{the signal-to-interference-plus-noise ratio}
\newacronym[]{msb}{MSB}{most-significative bit}
\newacronym[]{bcjr}{BCJR}{Bahl--Cocke--Jelinek--Raviv}
\newacronym[\glsshortpluralkey=SEDs,\glslongpluralkey=squared Euclidean distances]{sed}{SED}{squared Euclidean distance}
\newacronym[\glsshortpluralkey=EDs,\glslongpluralkey=Euclidean distances]{ed}{ED}{Euclidean distance}
\newacronym[\glsshortpluralkey=MEDs,\glslongpluralkey=minimum Euclidean distances]{med}{MED}{minimum Euclidean distance}
\newacronym[]{core}{CoRe}{constellation rearrangement}
\newacronym[]{msd}{MSD}{multistage decoding}
\newacronym[]{pdl}{PDL}{parallel decoding of the individual levels}
\newacronym[\glsshortpluralkey=GCs,\glslongpluralkey=Gray codes]{gc}{GC}{Gray code}
\newacronym[]{brgc}{BRGC}{binary-reflected Gray code}
\newacronym[]{nbc}{NBC}{natural binary code}
\newacronym[]{fbc}{FBC}{folded-binary code}
\newacronym[]{bsgc}{BSGC}{binary semi-Gray code}
\newacronym[]{msp}{MSP}{modified set-partitioning}
\newacronym[]{ssp}{SSP}{semi set-partitioning}
\newacronym[]{fhd}{FHD}{free Hamming distance}
\newacronym[]{mfhd}{MFHD}{maximum free Hamming distance}
\newacronym[]{ods}{ODS}{optimal distance spectrum}
\newacronym[]{iud}{i.u.d.}{independent and uniformly distributed}
\newacronym[]{ud}{u.d.}{uniformly distributed}
\newacronym[]{iid}{i.i.d.}{independent, identically distributed}
\newacronym[]{ami}{AMI}{accumulated mutual information}
\newacronym[]{bico}{BICO}{binary-input continuous-output}
\newacronym[]{gh}{GH}{Gauss--Hermite}

\newacronym[\glsshortpluralkey=BSs,\glslongpluralkey=base-stations]{bs}{BS}{base-station}
\newacronym[\glsshortpluralkey=MSs,\glslongpluralkey=mobile-stations]{ms}{MS}{mobile-stations}

\newacronym[]{phy}{PHY}{physical layer} 
\newacronym[]{rlc}{RLC}{Radio-Link control} 
\newacronym[]{ran}{RAN}{Radio Access Network} 
\newacronym[]{llc}{LLC}{logical link control} 
\newacronym[]{tcp}{TCP}{transmission control protocol} 
\newacronym[]{mac}{MAC}{media access control} 
\newacronym[]{fft}{FFT}{fast Fourier transform} 
\newacronym[]{ft}{FT}{Fourrier transform}
\newacronym[]{cf}{CF}{characteristic function} 
\newacronym[]{mgf}{MGF}{moment generating function} 
\newacronym[]{ee}{EE}{energy efficiency} 
\newacronym[]{eb}{EB}{energy per bit}
\newacronym[]{kkt}{KKT}{Karush--Kuhn--Tucker} 
\newacronym[]{mcs}{MCS}{modulation/coding scheme} 
\newacronym[]{fec}{FEC}{forward error correction}
\newacronym[]{arq}{ARQ}{automatic repeat request}
\newacronym[]{harq}{HARQ}{hybrid ARQ}
\newacronym[]{tarq}{TARQ}{truncated HARQ}
\newacronym[]{ir}{IR}{incremental redundancy}
\newacronym[]{rpr}{RR}{repetition redundancy}
\newacronym[]{rrharq}{RR-HARQ}{repetition redundancy HARQ}
\newacronym[]{irharq}{IR-HARQ}{incremental redundancy HARQ}
\newacronym[]{ack}{ACK}{positive acknowledgment}
\newacronym[]{nack}{NACK}{negative acknowledgment}
\newacronym[]{hol}{HoL}{head of the line}
\newacronym[]{crc}{CRC}{cyclic redundancy check}
\newacronym[]{dp}{DP}{dynamic programming}
\newacronym[]{gp}{GP}{geometric programming}
\newacronym[]{per}{PER}{packet error rate}
\newacronym[]{ber}{BER}{bit error rate}
\newacronym[]{op}{OP}{outage probability}
\newacronym[]{spa}{SPA}{saddle-point approximation}
\newacronym[]{mrc}{MRC}{maximum ratio combining}
\newacronym[]{mdp}{MDP}{Markov decision process}
\newacronym[]{lp}{LP}{linear programming}
\newacronym[]{pomdp}{POMDP}{Partially observable MDP}
\newacronym[]{psimdp}{PSI-MDP}{Partial State Information Markov Decision Process}
\newacronym[]{scpp}{SCPP}{stochastic shortest path problem}

\newacronym[]{mm}{MM-HARQ}{multi-message HARQ}
\newacronym[]{xp}{XP-HARQ}{cross-packet HARQ}
\newacronym[]{ts}{TS}{time-sharing}
\newacronym[]{sc}{SC}{superposition coding}
\newacronym[]{sbrq}{SBRQ}{systematic backward retransmission}
\newacronym[]{brq}{BRQ}{backward retransmission}
\newacronym[]{lharq}{L-HARQ}{layer-coded HARQ}
\newacronym[]{anlharq}{AoN-HARQ}{all-or-none L-HARQ}
\newacronym[]{vlharq}{VL-HARQ}{variable-length HARQ}

\newacronym[]{pp}{PPP}{point process}
\newacronym[]{ppp}{PPP}{Poisson point process}
\newacronym[]{pgfl}{PGFL}{Poisson point process}

\usepackage{tikz,pgfplots}
\pgfplotsset{width=3.4in,height=2.3in}
\usetikzlibrary{positioning,shadows,shapes,fit,arrows,backgrounds}  

\newcounter{chem}
\newcounter{temp}
\newenvironment{Sequation}{%
  \setcounter{temp}{\value{equation}}%
  \setcounter{equation}{\value{chem}}%
}{%
  \setcounter{chem}{\value{equation}}%
  \setcounter{equation}{\value{temp}}%
}

\usepackage{epsf}           
\usepackage{graphicx}       
\usepackage{epsfig}         
\usepackage{pstricks}
\usepackage{pstricks-add}
\usepackage{pst-plot}
\usepackage{dsfont}
\newcounter{defcounter}
\usepackage{balance}

\newtheorem{example}{Example}

\newcommand{\siz}{0.85}  
\newcommand{\sizf}{0.9}   

\usepackage{soul}  

\begin{document}


\title{AMC and HARQ: \\How to Increase the Throughput}

\author{Mohammed Jabi, Leszek Szczecinski, Mustapha Benjillali, Abdellatif Benyouss, and Benoit Pelletier
\thanks{%
M. Jabi was with 
INRS, Montreal, Canada; he is now with Nokia Poland. [e-mail: jabi.mohamed@gmail.com].}
\thanks{%
L. Szczecinski, and A. Benyouss are with 
INRS, Montreal, Canada. [e-mail: leszek@emt.inrs.ca,benyoussabdellatif@gmail.com].}
\thanks{%
M. Benjillali is with the Communication Systems Department, INPT, Rabat, Morocco. [e-mail: benjillali@ieee.org]. }
\thanks{%
Benoit Pelletier is with InterDigital Canada, Lt\'ee., Montreal, Canada. [e-mail: Benoit.Pelletier@interdigital.com]}%
}%

\maketitle
\thispagestyle{empty}

\begin{abstract}
In this work, we consider transmissions over block fading channels and assume that adaptive modulation and coding (AMC) and hybrid automatic repeat request (HARQ) are implemented. Knowing that in high signal-to-noise ratio, the conventional combination of HARQ with AMC is counterproductive from the throughput point of view, we adopt the so-called layer-coded HARQ (L-HARQ). L-HARQ allows consecutive packets to share the channel and preserves a great degree of separation between AMC and HARQ; this makes the encoding and decoding very simple and allows us to use the available/optimized codes. Numerical examples shown in the paper indicate that L-HARQ can provide significant throughput gains compared to the conventional HARQ. The L-HARQ is also implemented using turbo codes indicating that the throughput gains also materialize in practice.
\end{abstract}


\section{Introduction}\label{Sec:Intro}

\IEEEPARstart{I}{n} this work, we are interested in increasing the throughput of the \gls{phy} when the coded information is transmitted using equal-length channel blocks which are subject to independent fading. We are motivated by two main results. First, by \cite{Sassioui17}, which demonstrates that using the conventional \gls{harq} when \gls{amc} is adopted decreases the throughput at high \gls{snr}. The second result is due to \cite{Popovski14,Jabi17b} which proposes \gls{lharq} -- a simple encoding/decoding  scheme tailored to improve the throughput of \gls{harq}. However,  \gls{lharq} was studied when the instantaneous \gls{csi} was not available at the transmitter; \ie without the \gls{amc}. In this work we propose to leverage  the knowledge of instantaneous \gls{csi} and incorporate it in the coding scheme of \gls{lharq}.

\subsection{Background}

\gls{amc} is adopted for communications over time-varying channels to guarantee an efficient use of channel resources. It relies on a feedback channel over which the receiver informs the transmitter about the ``best'' transmission rate to use. This rate is evaluated by the receiver based on the estimated \gls{csi}. Despite this rate adaptation, transmission errors are unavoidable in practice, and they are handled by the retransmission protocol, \gls{harq}. 

\gls{harq} also uses  feedback channel: one-bit messages inform the transmitter about the decoding success (\gls{ack}) or failure (\gls{nack}). After each \gls{nack}, the transmitter starts a new \gls{harq} \emph{round} (\ie a retransmission) which conveys additional information necessary to decode the packet; in other words, \gls{harq} encodes the packet \emph{across} the transmission rounds. The \gls{harq} \emph{cycle}, defined as the sequence of transmission rounds of the same packet, terminates if the packet is correctly received (as indicated by \gls{ack}). If the number of rounds is limited, \gls{harq} is said to be \emph{truncated}; then, the \gls{nack} in the final round indicates a \emph{packet loss}.


Both, \gls{amc} and \gls{harq} may be considered as parts of  \gls{phy},  and their interaction has been addressed vastly in the  literature. For instance, \cite{Liu04,Zheng05,Kim08b,Ramis11,Zhang13,Kang10} analyzed  the throughput, while \cite{Le06,Wang07,Le07,Femenias09} focused on the delay due to \gls{harq}. In all these works, a constraint on the probability of packet loss was imposed; by doing so, the value of  \gls{harq} was highlighted as, indeed, \gls{harq} efficiently decreases the probability of having packets lost at the end of the cycle. This is different from our work because we ignore the packet loss and only sheer throughput is considered; the rationale for such an approach is discussed in \secref{Sec:upper.layers}. 

To meaningfully compare different \gls{amc}/\gls{harq} strategies, it is important to assume that the ressources used by the \gls{phy} are fixed.\footnote{First, this is because the block fading model implicitly assumes that the duration of all transmissions is the same. More importantly, from a methodological point of view, the transmission with variable \gls{phy} resources is more relevant in the context of multi-user communications, which also implies some form of resource management; then, drawing useful conclusions may be difficult analyzing solely the \gls{phy}. Thus, fixing the resources of the point-to-point \gls{phy}, we are able to isolate \gls{harq} from the external factors; such as a multi-user scheduling \cite{Poggioni10,Carrasco2013} or a cooperative transmission \cite{Harsini11}; which may be combined with \gls{harq} but are difficult to characterize.} This goal can be attained by encoding one packet over the whole channel block in each round; such \gls{harq} is said to be \emph{conventional}, and was often used as a basis for theoretical analysis, \eg \cite{Caire01}. 

However, as demonstrated in \cite{Sassioui17}, this conventional approach can actually be detrimental to the throughput; that is, \gls{amc} alone is better than \gls{amc} combined with the conventional \gls{harq}. This surprising, at first sight, result is caused by the fact that the conventional \gls{harq} is not fully adaptive: it uses the \gls{csi} only in the first transmission round \cite{Sassioui16} of a given packet; reusing next the entire channel block for each round causes a waste of resources.\footnote{We will comment on this in \exref{Ex:QAM.Rayleigh}, but note here that a somehow similar behavior of the throughput is also observed in \gls{harq} transmissions without \gls{csi}: for high \gls{snr}, the throughput of \gls{harq} improves only negligibly \cite{Larsson14,Jabi15b} when the number of rounds increases.}

Indeed, the literature already recognizes that, to improve the throughput, the coding across the \gls{harq} rounds must be modified.  The most relevant solutions may be classified as {\it{i)}}~a \emph{multi-packet coding} \cite{Popovski14,Jabi15b,Jabi17a,Jabi17b}, where many packets with variable contents are jointly encoded into fixed-length codewords which then use the fixed resources (channel blocks) or, as {\it{ii)}}~a \emph{variable-length coding}\cite{Cheng03,Uhlemann03,Visotsky05,Pfletschinger10,Szczecinski13,Jabi16}, where rather the codewords length varies throughout the \gls{harq} rounds and the packet content is fixed.\footnote{We also note that similar ideas are discussed in \cite{Verdu10}, where the \emph{fixed-to-variable} (here, \emph{variable-length}) and \emph{variable-to-fixed} (here, \emph{multi-packet}) coding strategies are defined. In fact, the \gls{lte} standard enables the variable-length coding \cite[Ch.~12.1]{Dahlman14_book} varying the number of the so-called resource blocks \cite[Ch.~9.1]{Dahlman14_book} on a per \gls{harq}-round basis.}


We focus this work on the adaptive multi-packet coding whose advantage over the relatively well-studied variable-length coding will be discussed in \secref{Sec:VLHARQ}.


\subsection{Contribution and Organization}

The main difficulties of the multi packet coding are {\it{i)}}~the practicality of joint encoding/decoding of many packets, and {\it{ii)}}~the rate adaptation of this joint coding. This was already apparent in the case of multi-packet coding when the instantaneous \gls{csi} was not available at the transmitter \cite[Sec. V]{Jabi17a}.

To make the multi-packet approach practical, we will use \gls{lharq}, studied in \cite{Popovski14,Jabi17b} and implemented via a  two-step (or, layered) approach, where the joint coding is implemented in two independent steps: the binary packet mixing is followed by the conventional channel coding. From the theoretical perspective, when compared to a more general joint coding/decoding scheme \cite{Jabi17a,Trillingsgaard17}, \gls{lharq} does not impose any throughput penalty \cite[Th. 3]{Trillingsgaard17}. It is also more practical as it can be used with commercially available encoders/decoders \cite{Jabi17b}. The remaining issue is how to use \gls{lharq} with \gls{amc}, that is, how to exploit the \gls{csi} in all transmission rounds and yet maintain the simple layered-coding strategy.

The main contribution of this work lies, therefore, in the generalization of \gls{lharq} \cite{Jabi17b,Popovski14} to take advantage of the \gls{csi} in all rounds of \gls{harq}. The proposed encoding scheme yields a fully adaptive \gls{harq}: like in \cite{Jabi17b,Popovski14}, the \gls{csi} observed in the previous rounds of the same packet is exploited but, unlike \cite{Jabi17b,Popovski14}, each round exploits as well the instantaneous (possibly outdated) \gls{csi}.

The question which needs to be answered in this context is: how to adjust the coding rates using the rich information about the \gls{csi}? We formulate the optimization problem and propose a framework to solve it. Concluding on the excessive numerical complexity of the latter, we present heuristic rate-adaptation policies inspired by the optimal adaptation policy used in the \gls{amc} and by the results of \cite{Jabi17b}. The proposed solution leads to a local adaptation with easily adjustable parameters and does not require the knowledge of the entire model of the relationship between the different \gls{harq} rounds, the \gls{amc}, and the fading model.

To the best of our knowledge, the literature does not provide such a solution to adapt \emph{simultaneously} the \gls{amc} and \gls{harq}, while keeping the channel resources fixed. The most practical element is that the adaptation we propose relies on the error-rate curves of the commercial encoders/decoders that can be easily acquired. This is critical from the implementability perspective. In fact, rare are works which include the practical encoders/decoders, mainly because the decoding error-rate curves after many \gls{harq} rounds are difficult to obtain and to use; for example, a significant part of  \cite{Visotsky05} (which implements a variable-length \gls{harq}) is dedicated to the issue of approximating these curves and the results are limited to three \gls{harq} rounds.

The rest of the paper is organized as follows. In~\secref{Sec:Model} we introduce the adopted models while the principle of the  proposed \gls{lharq} is explained in~\secref{sec:lharq.principle}. The optimization issues are discussed in \secref{Sec:Rate.adaptation} and \secref{Sec:simple.rate.adaptation}. \secref{Sec:examples} compares the proposed \gls{lharq} to the alternative strategies using examples of {\it{i)}}~the idealized encoders/decoders, and {\it{ii)}}~ the practical turbo-codes. The conclusions are drawn in~\secref{Sec:Conclusions}.

\section{Model}\label{Sec:Model}

\begin{figure}[tb]
\begin{center}
\scalebox{.7}{

\pgfdeclarelayer{background}
\pgfdeclarelayer{foreground}
\pgfsetlayers{background,main,foreground}


\tikzstyle{form1} = [draw, minimum width=2cm, text width=1.8cm, fill=gray!15, 
  text centered,  minimum height=.9cm]

\tikzstyle{form2} = [draw=none, minimum width=3.4cm, text width=1.5cm, fill=none, 
  text centered,  minimum height=0cm]    

\tikzstyle{form3} = [draw,dashed, minimum width=1.8cm, text width=1.8cm, fill=red!10, 
  text centered,  minimum height=1.8cm]
  
  \tikzstyle{form4} = [draw=none, minimum width=0cm, text width=0cm, fill=none, 
  text centered,  minimum height=.2cm] 
  \tikzstyle{form5} = [draw=none, minimum width=0cm, text width=0cm, fill=black, 
  text centered,  minimum height=.6cm] 
    
    \tikzstyle{elip}=[draw,ellipse, minimum width=.2cm, text width=0cm, fill=none, 
  text centered,  minimum height=1cm]
  
    \tikzstyle{elip2}=[draw,ellipse, minimum width=.1cm, text width=0cm, fill=none, 
  text centered,  minimum height=.6cm]

\begin{tikzpicture}[trim left=-5cm]

    \node(Channel)[form1] {Channel};
    
     \path (Channel.west)+(-2.7,0)  node[form1](Enc) {Encoder};
     \path (Channel.east)+(2.7,0)  node[form1](Dec) {Decoder};
       \path (Enc.north)+(0,1.6)  node[form1](AMC_cont1) {AMC Controller};
        \path (AMC_cont1.north)+(0,0.45)  node[form1](HARQ_cont1) {HARQ Controller};
         \path (HARQ_cont1.north)+(0,1.6)  node[form1](ARQ_cont1) {ARQ Controller};
          \path (HARQ_cont1.west)+(-0.8,1)  node(P1) {};
      \path (Dec.north)+(0,1.6)  node[form1](AMC_cont2) {AMC Controller};
       \path (AMC_cont2.north)+(0,0.45)  node[form1](HARQ_cont2) {HARQ Controller};
         \path (HARQ_cont2.north)+(0,1.6)  node[form1](ARQ_cont2) {ARQ Controller};
       \path (HARQ_cont2.east)+(0.8,1)  node(P2) {};
       \path (Enc.west)+(-1,-0.2)  node(Buff1) {};
       \path (Enc.south)+(0,-1)  node(Tx) {Transmitter};
        \path (Dec.south)+(0,-1)  node(Rx) {Receiver};
        
         \path(ARQ_cont1.north) +(0,0.6) node(base1){};
       \path(ARQ_cont1.north) +(1,0.6) node(Buff11){};
       \path(ARQ_cont1.north) +(-1,0.6) node(Buff12){};
       \path(ARQ_cont1.north) +(1,0.8) node(Buff13){};
       \path(ARQ_cont1.north) +(-1,0.8) node(Buff14){};
       \path(ARQ_cont1.north) +(1,1) node(Buff15){};
       \path(ARQ_cont1.north) +(-1,1) node(Buff16){};
       \path(ARQ_cont1.north) +(1,1.6) node(Buff17){};
       \path(ARQ_cont1.north) +(-1,1.6) node(Buff18){};

       \path(ARQ_cont2.north) +(0,0.6) node(base2){};
       \path(ARQ_cont2.north) +(1,0.6) node(Buff21){};
       \path(ARQ_cont2.north) +(-1,0.6) node(Buff22){};
       \path(ARQ_cont2.north) +(1,0.8) node(Buff23){};
       \path(ARQ_cont2.north) +(-1,0.8) node(Buff24){};
       \path(ARQ_cont2.north) +(1,1) node(Buff25){};
       \path(ARQ_cont2.north) +(-1,1) node(Buff26){};
       \path(ARQ_cont2.north) +(1,1.6) node(Buff27){};
       \path(ARQ_cont2.north) +(-1,1.6) node(Buff28){};
%

 \path[draw] (Buff11.west)--(Buff12.east);
     \path[draw] (Buff13.west)--(Buff14.east);
     \path[draw] (Buff15.west)--(Buff16.east);
     \path[draw] (Buff11.west)--node[shift={(0.5,0)}]{}(Buff17.west);
     \path[draw] (Buff12.east)--(Buff18.east);
    \path[draw,<-,>=latex] (ARQ_cont1.north)--(base1.base);

    \path[draw] (Buff21.west)--(Buff22.east);
     \path[draw] (Buff23.west)--(Buff24.east);
     \path[draw] (Buff25.west)--(Buff26.east);
     \path[draw] (Buff21.west)--(Buff27.west);
     \path[draw] (Buff22.east)--(Buff28.east);
    \path[draw,->,>=latex] (ARQ_cont2.north)--(base2.base);

 \path[draw,->,>=latex] (Enc.east)--
 node[near end,above]{$\bx[n]$}
 node[below]{$\R(\SNR)$}
 (Channel.west);
  \path[draw,->,>=latex] (Channel.east)--
  node[near end,above]{$\by[n]$}
  node[below]{$\tilde{\SNR}$}
  (Dec.west);

     \path[draw,->,>=latex] (AMC_cont1.south)--
     node[near end,left]{$\mfm[n]$}
     (Enc.north);
     \path[draw,<-,>=latex] (AMC_cont2.south)--
     node[near end,right]{$\hat{\mfm}[n]$}
     (Dec.north);
     \path[draw,<-,>=latex] (ARQ_cont2.south)--(HARQ_cont2.north);
      \path[draw,->,>=latex] (ARQ_cont1.south)--(HARQ_cont1.north);
     
      
      \path[draw,<-,>=latex,dashed] (AMC_cont1.east)--node[shift={(0,.3)}]{$\CSI=\SNR$, $\ell$}(AMC_cont2.west);
      \path[draw,<-,>=latex,dashed] (ARQ_cont1.east)--node[shift={(0,.3)}]{LLC-level ACK/NACK}(ARQ_cont2.west);
       \path[draw,<-,>=latex,dashed] (HARQ_cont1.east)--node[shift={(0,.3)}]{PHY-level ACK/NACK}(HARQ_cont2.west);

      \path[draw,>=latex,dashed] (P1.east)--(P2.west);
       \path (P1.north)+(0,0.3)  node(gamma0) {LLC};
        \path (P1.south)+(0,-0.3)  node(gamma0) {PHY};
        \path (P2.north)+(0,0.3)  node(gamma0) {LLC};
        \path (P2.south)+(0,-0.3)  node(gamma0) {PHY};


\end{tikzpicture}}
\end{center}
\caption{Model of the transmission of the packet $\mfm[n]$:  \gls{amc} chooses the rate based on the \gls{csi}, $\CSI$, provided by the receiver before the transmission begins; here, the \gls{csi} is the same as the \gls{snr} and with a finite number of rates, it is enough to transmit the index of the rate, $\ell$, and the transmission rate is then given by $\R(\SNR)=\R_{[\ell]}$. The \gls{llc} layer implements the \gls{arq} which ensures the error-free transmission and makes the throughput of the \gls{phy} the unique criterion for comparison, see \secref{Sec:upper.layers}.}\label{fig:HARQ.AMC.model}
\end{figure}
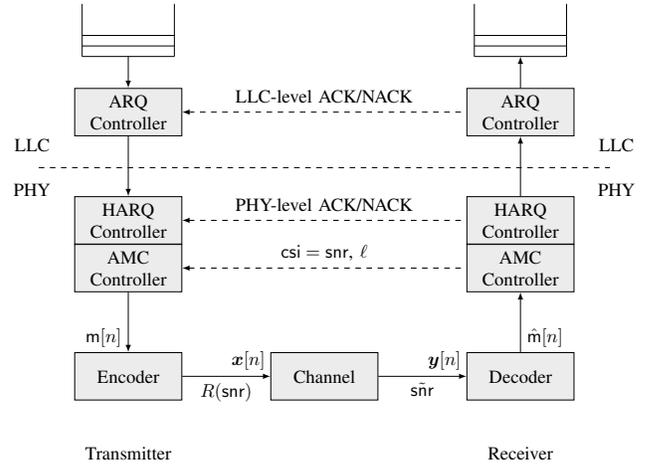

\subsection{AMC}\label{Sec:PHY}

In a point-to-point transmission over a block fading channel, illustrated in \figref{fig:HARQ.AMC.model}, we assume that the size, $\Ns$, of the transmission blocks, $\bx[n]$, does not change with their index $n$;  the transmitter encodes the packet $\mfm[n] \in\set{0,1}^{R  \Ns}$ into a codeword $\bx[n]=\Phi\big[\mfm[n]\big]\in\X^{\Ns}$, where $\Phi[\cd]$ is the encoder, $\R$ -- the coding rate, and $\X$ -- the constellation set of size $M=|\X|$. 

We assume that the choice of the rate is done at the receiver using the estimated \gls{csi}, $\CSI$, that is, $\R=\R(\CSI)\in\mcR$, where $\mcR$ is the set of available rates. To simplify the examples, we will consider frequency non-selective channel, where the \gls{csi} is uniquely represented by the \gls{snr}, $\SNR$. 

While, for theoretical considerations we may sometimes assume a continuous-valued set $\mcR=\Real_{+}$ is available, in practice, the set of available rates is discrete, \ie $\mcR=\set{\R_{[1]},\ld, \R_{[L]}}$ and then it is enough to transmit the \emph{index} $\ell\in\set{1,\ld, L}$ of the chosen rate, that is $\R(\SNR)=\R_{[\ell]}$; here, $L$ denotes the number of available rates.


In all examples, $\X$ will be a $M$-ary \gls{qam} constellation (with $M=4$ or $M=16$);  the empirical error-rate curves of the practical encoders/decoders will be obtained using the turbo encoding/decoding. We emphasize that, independently from how the coded modulation is implemented, there is always a maximum rate which cannot be exceeded; in our case, this limitation is expressed as $\R_{[L]}<\log_2(M)$. 

Assuming that packets are always available at the transmitter (the buffer is saturated), the codeword $\bx[n]$ is transmitted in the $n$th block, and the receiver observes the signal
\begin{align}\label{y.x.z}
\by[n]=\sqrt{\tilde{\SNR}[n]} \bx[n]+\bz[n],
\end{align}
where $\bz[n]$ is a vector of realizations of zero mean, unit-variance, complex Gaussian variables, modeling the noise. Since $\bx[n]$ are (realizations of) random variables uniformly distributed over $\X$, with appropriate normalization of the latter, $\tilde{\SNR}[n]$ is the \gls{snr} at the receiver experienced at the time the transmission is carried out. 

In general, $\tilde{\SNR} \neq \SNR$, and this \gls{snr} ``mismatch'' is caused by the estimation errors and/or by the delay between the time the \gls{snr}, $\SNR[n]$, is estimated and the time the transmission is carried out with the experienced \gls{snr}, $\tilde{\SNR}[n]$.\footnote{Although the time index of both \gls{snr}s refers to the same $n$th channel block, the estimation must be carried out before the transmission takes place; it has to be done sufficiently early to leave a time for a feedback of the index $\ell$ and to allow the transmitter to encode/modulate the packet. Thus, there is always a delay between the moments of estimation and transmission.}


The receiver tries to decode the transmitted packet $\mfm[n]$ using the channel outcome
\begin{align}\label{hat.m}
\hat{\mfm}[n]=\tr{DEC}\big[\by[n]\big].
\end{align}

The decoding errors \mbox{$\Err[n]=\set{\hat{\mfm}[n]\neq\mfm[n]}$} that occur due to different, simultaneously occurring events (such as \eg atypical noise, interference, or fading) are characterized by a  \gls{per} function 
\begin{align}
\label{PER.func}
\PER(\SNR;\R)\triangleq \PR{\Err|\SNRrv=\SNR, \R},
\end{align}
which, for a given $\R$, decreases monotonically with $\SNR$. This function captures the behavior of the entire receiver from the point of view of the transmitter and includes the effect of the decoding, channel estimation, and synchronization. It also captures the effect of delayed/imperfect \gls{csi} as follows
\begin{align}\label{PER.func.Ex}
\PER(\SNR;\R)&=\Ex_{\tilde{\SNRrv}}\left[\PER^\tnr{rx}(\tilde{\SNRrv};\R)| \SNRrv=\SNR\right]
\end{align}
where
\begin{align}\label{}
\label{PER.Rx.func}
\PER^\tnr{rx}(\tilde{\SNR};\R)\triangleq \PR{\Err|\tilde{\SNRrv}=\tilde{\SNR}, \R},
\end{align}
is the \gls{per} function of the receiver and depends on the \gls{snr}, $\tilde{\SNR}$, actually observed and experienced during the transmission.

In practice, it is not necessary to estimate/measure the entire \gls{per} function and it is sufficient to find the \gls{snr} interval limits $\gamma_{[\ell]}$ which satisfy a constraint on \gls{per} when transmitting with rate $\R_{[\ell]}$ \cite[Sec.~III]{Sassioui17}, \ie
\begin{align}\label{gamma.ell.eps}
\gamma_{[\ell]}=\min_{\SNR \in \Real^{+}} \set{\SNR: \PER(\SNR;\R_{[\ell]})\le\epsilon},
\end{align}
where $\epsilon \in \Real^{+}$ denotes a \gls{per} constraint. This approach can be related to the calibration of the \gls{lte} receivers which, observing the \gls{csi} (here the \gls{snr}, $\SNR$), should report the largest rate $\R_{[\ell]}\in \mcR$ for which the inequality condition in~\eqref{gamma.ell.eps} is satisfied with $\epsilon=10^{-1}$ \cite[Sec.~7.2]{3GPP_TS_36.213}. That is, the calibration process in \gls{lte} implicitly solves \eqref{gamma.ell.eps}. The coding adaptation for \gls{harq} we will propose, exploits  the \gls{per} curves (or the \gls{snr} intervals) and thus preserves the legacy and adaptation simplicity of current systems.


\subsection{Channel Model}
We will model $\SNR[n]$ as realizations of \gls{iid} random variables $\SNRrv[n]$, which is appropriate if the transmission of the channel blocks $\bx[n]$ is well separated in time.\footnote{In the \gls{lte}, the channel blocks are attributed to many users in a time-interleaved manner. So, here, the blocks $\bx[n]$ correspond to a particular user but are attributed in non-adjacent time instants. This time-separation allows to account for the round-trip delay due to the propagation, and the processing at the receiver and the transmitter.}

The derivations will be done in abstraction of a particular fading type, but in the numerical examples, we assume Rayleigh fading, \ie $\SNRrv[n]$ is drawn from an exponential distribution
\begin{align}\label{pdf.SNR}
\pdf_{\SNRrv}(\SNR)=1/\SNRav\exp(-\SNR/\SNRav),
\end{align}
where $\SNRav$ is the average \gls{snr}, and we have removed the time-indexing $[n]$ that is irrelevant with the \gls{iid} modeling of the \glspl{snr}.

The variables $\SNRrv[n]$ and $\tilde{\SNRrv}[n]$ are, in general, mutually dependent. Again, for the sake of numerical examples, we assume that the \gls{snr} experienced during the transmission, $\tilde{\SNRrv}$, is a delayed version of the \gls{snr} estimated at the receiver, $\SNRrv$,\footnote{As we said before, $\SNR$ is estimated first by the receiver, used by the transmitter to adapt the rate $\R(\SNR)$, and the transmission is finally carried out with the \gls{snr} $\tilde{\SNR}$.}  and their  joint \gls{pdf} is given by  \cite{Alouini00,Levorato09}
\begin{align} \label{eq:cond.probablity}
\pdf_{\SNRrv,\tilde{\SNRrv}}(\SNR,\tilde{\SNR})=&\frac{1}{(1-\corSNR)\SNRav} ~I_0\!\!\left(\!\frac{2\sqrt{\corSNR \SNR~ \tilde{\SNR}}}{(1-\corSNR)\SNRav} \right) \nonumber \\ 
&\cd\exp \left(-\frac{\tilde{\SNR}+\SNR}{(1-\corSNR)\SNRav}\right),
\end{align}
where $I_{0}$ is the zero-order modified Bessel function of the first kind, and  $\corSNR$ is the correlation factor
\begin{align}
\corSNR=J_0^2(2\pi \fD \tau).
\end{align}
Here, $\tau$ is the time difference between the instants of estimation and transmission, $\fD$ is the Doppler frequency, and $J_0$ is the zero-order Bessel function of the first kind. Essentially the same relationship will be obtained if we assume that the difference between $\tilde{\SNR}$ and $\SNR$ is due to channel estimation errors \cite[eq. (20)]{Hu13}.

\begin{example}[Threshold decoding]\label{ex:threshold.Doppler}
Assume that 
the receiver's \gls{per} function is binary
\begin{align}\label{}
\PER^\tnr{rx}(\tilde{\SNR};\R_{[\ell]})=\IND{\tilde{\SNRrv}<\gamma_{[\ell]}^{\mr{th}}},
\end{align}
where the indicator function $\IND{a}=1$ if  $a$ is true, and $\IND{a}=0$ otherwise; and $\gamma_{[\ell]}^{\mr{th}}$ is uniquely defined by the transmission rate $\R_{[\ell]}$.

If the joint distribution of $\SNRrv$ and $\tilde{\SNRrv}$ is defined by \eqref{eq:cond.probablity},  the \gls{per} function \eqref{PER.func} is calculated using \eqref{PER.func.Ex} as 
\begin{align}\label{eq:per.example}
\PER(\SNR;\R_{[\ell]})
&=\PR{\tilde{\SNRrv}<\gamma_{[\ell]}^{\mr{th}}| \SNRrv=\SNR,\R_{[\ell]}}\\
\label{per.Q1}
&=
Q_{1}\left( \sqrt{\frac{2\corSNR\SNR}{(1-\corSNR)\SNRav }}, \sqrt{\frac{2 \gamma_{[\ell]}^{\mr{th}}}{(1-\corSNR)\SNRav }}   \right),
\end{align}
where we used the form of the (complementary) \gls{cdf} of a non-central chi-square distribution \cite[Sec. 2.3]{Proakis08_Book} with $Q_{m}(\cd,\cd)$ being the generalized Marcum $Q$-function.
\end{example}

\subsection{Throughput-Optimal Rate Adaptation}\label{Sec:Throughput}

The throughput is defined as the long-term average number of successfully received bits per transmitted symbol, or channel use, ($\bpcuval$) and, since we use constant-length blocks, we can write it as
\begin{align}\label{eta.def.time}
\eta=\lim_{N\rightarrow\infty}\frac{1}{N}\sum_{n=1}^N \mfR[n],
\end{align}
where $\mfR[n]$ denotes the number of decoded bits normalized by the block length, $\Ns$, \ie $\mfR[n]\in\set{0, \R(\SNR[n])}$. It is also called a  \emph{reward} in the $n$-th block.

The mean reward (expectation taken with respect to the decoding error events, $\Err$) is given by \mbox{$\R \big(1-\PER(\SNR;\R)\big)$} and, since we model $\SNR[n]$ by \gls{iid} random variables,  the time-average from \eqref{eta.def.time} may be replaced by the expectation \cite{Hu13,Sassioui17}
\begin{align}
\label{eta.amc.1}
\eta^{\amc}
=&\Ex_{\SNRrv}\bigg[\R(\SNRrv)\cd \Big(1-\PER\big(\SNRrv;\R(\SNRrv)\big)\Big)\bigg].
\end{align}

To find the rate adaptation function $\R(\SNR)$ which maximizes $\eta^{\amc}$, we can enter with the maximization under the expectation operator in \eqref{eta.amc.1} and the throughput-optimal rate adaptation function $\R(\SNR)$ of the \gls{amc} is then found by solving the following one-dimensional optimization problem (for each value of $\SNR$)
\begin{align}\label{eq:optimal/policy.K.1}
\R(\SNR)=\argmax_{\R\in\mcR}~\R\big(1-\PER(\SNR;\R)\big).
\end{align}

In practice, $\R(\SNR)$ is monotonically increasing with $\SNR$, which is in line with our intuition: for high \gls{snr} (or, more generally, for high-quality \gls{csi}), a larger transmission rate is used and a larger throughput is obtained. 

So, for the discrete rates we identify the \emph{decision regions} of the \gls{snr}, $[\gamma_{[\ell]}, \gamma_{[\ell+1]})$ such that the rate adaptation is defined as 
\begin{align}\label{adapt.func}
\R_{[\ell]}=\R(\SNR) \iff \SNR\in[\gamma_{[\ell]}, \gamma_{[\ell+1]}),
\end{align}
where $\gamma_{[1]}\triangleq0$, $\gamma_{[L+1]}\triangleq\infty$.

We mention briefly that the solutions of \eqref{eq:optimal/policy.K.1} and \eqref{gamma.ell.eps} do not, in general, produce the same decision regions. The choice of the one or the other criterion to define the \gls{snr} limits $\gamma_{[\ell]}$ is an implementation issue.

We emphasize here that, while the throughput \eqref{eta.amc.1}  depends on the distribution $\pdf_{\SNRrv}(\SNR)$, the optimal adaptation function, $\R(\SNR)$ -- does not. In fact, the distribution $\pdf_{\SNRrv}(\SNR)$ is most likely not known in practice, so using the  rate-adaptation \eqref{eq:optimal/policy.K.1} or \eqref{adapt.func} which are oblivious to the knowledge of $\pdf_{\SNRrv}(\SNR)$ is a desirable feature. Of course, the \gls{per} function $\PER(\SNR;\R_{\ell})$ may depend on the probabilitic model of the relationship between $\SNRrv$ and $\tilde{\SNRrv}$ as we saw in \exref{ex:threshold.Doppler}; however, such a relationship will still be captured by $\gamma_{[\ell]}$, see \eqref{gamma.ell.eps}; these \gls{snr} interval limits are then optimal irrespectively of  the marginal distribution $\pdf_{\SNRrv}(\SNR)$ required to calculate the expectation in \eqref{eta.amc.1}.

\textbf{Fine-tuning the adaptation}

The rate adaptation defined by \eqref{eq:optimal/policy.K.1} is throughput-optimal, however, we may use $\R(\SNR)$ with additional heuristics targeting the error rates. Namely, we can artificially change the argument of the rate-adaptation function $\R(\cd)$ in \eqref{adapt.func} and use the rate
\begin{align}\label{delta.gamma}
\R\leftarrow \R( \SNR\Delta);
\end{align}
for $\Delta<1$, it provides a more ``conservative" adaptation: the same rates $\R_{[\ell]}$ will be  used in higher  \glspl{snr} (comparing to $\Delta=1$), which leads to a smaller probability of decoding error. On the other hand, the rate-adaptation is more ``aggressive'' with $\Delta>1$ which allows the transmitter to use higher rates at the cost of larger probability of error. This simple modification of the adaptation strategy which relies on a fine-tuning of one parameter, $\Delta$, will be exploited later. 

\textbf{Performance limit}

Since we do not consider power adaptation in this work, the throughput is always upper-bounded  by 
the ergodic capacity of the channel \cite{Alouini00,Caire01,Wu10}
\begin{align}
\eta&\leq \ov{C}\\
\label{C.erg}
\ov{C}&\triangleq \Ex_\SNRrv[ I(\SNRrv)],
\end{align} 
where  $I(\SNR)=\mfI(X;Y|\SNR)$ is the \gls{mi} between the random variables $X\in\X$ and $Y$  modeling respectively the channel input and output.

The limit  \eqref{C.erg} can be attained in the idealized case when {\it{i)}}~$\SNRrv =\tilde{\SNRrv}$, {\it{ii)}}~capacity achieving codes are used, {\it{iii)}}~the transmission rate set $\mcR$ is continuous, and {\it{iv)}}~the rates are adapted as $\R(\SNR)=I(\SNR)$ \cite{Alouini00}; then, the receiver's \gls{per} function \eqref{PER.Rx.func} is binary, \ie $\PER^\tnr{rx}(\tilde{\SNR};\R)=\IND{I(\tilde{\SNR})<\R}$; this is similar to the  assumption we made in \exref{ex:threshold.Doppler} with $\gamma_{[\ell]}^{\mr{th}}=I^{-1}(\R_{[\ell]})$.

\subsection{HARQ}\label{Sec:HARQ}

If retransmissions are allowed, the packet $\mfm \in\set{0,1}^{R_{[\ell]} \cd \Ns}$ is encoded into $\kmax$ subcodewords, $\bx_k=\Phi_k[\mfm]\in\X^{\Ns}$, where $\Phi_k[\cd], k=1,\ld, \kmax$, are the encoding functions, and $\kmax$ is the maximum number of transmissions for each packet. We consider here \gls{irharq}, that is, all the subcodewords $\bx_k$ are complementary (punctured) versions of a mother codeword, $\bx_\tr{o}=[\bx_1,\ld, \bx_K]$. 

Then, each  round carries a different subcodeword $\bx_k$, and the transmission outcome of the $k$-th round is given by
\begin{align}\label{y.x.z.harq}
\by_{k}=\sqrt{\tilde{\SNR}_{k}} \bx_{k}+\bz_{k},\quad k=1,\ld, \kmax,
\end{align}
where we eliminate the time-index $n$ from \eqref{y.x.z} and rather use the ``packet-centric'' notation, indexing the \gls{harq} rounds with $k$;  thus, $\tilde{\SNR}_{k}$ is the \gls{snr} at the receiver in the $k$-th \gls{harq} round. The important point in the ``conventional'' \gls{irharq} is that the rate is chosen in the first round, \ie $\R_{[\ell]}=\R^{\tnr{harq}}(\SNR_1)$.


%

The question which should be answered now is: how to find the \gls{harq} throughput-optimal rate adaptation $\R^{\tnr{harq}}(\SNR_1)$? 


Since \gls{harq} introduces memory, the optimal rate-adaptation function $\R^{\tnr{harq}}(\SNR)$  depends on the distribution of the \gls{snr} and is quite difficult to find.\footnote{In particular, because the decision regions are not simply defined by the intervals \eqref{adapt.func}, \cite{Sassioui17}.} This clashes with the simplicity of the \gls{amc} rate-adaptation \eqref{eq:optimal/policy.K.1} which does not depend on the distribution $\pdf_{\SNRrv}(\SNR)$, see comments after \eqref{adapt.func}. 

Fine-tuning strategy \eqref{delta.gamma} might be used as a practical alternative. However, optimizing/fine-tuning $\R^\tnr{harq}(\SNR_1)$ is not always worth the effort: we know from \cite[Prop.~4]{Sassioui17} that, using \gls{harq} on top of the \gls{amc}, actually penalizes the throughput if we consider the region of high \gls{snr}. This happens, independently of how we choose the rate adaptation function $\R^\tnr{harq}(\SNR_1)$, because \gls{harq} adapts the rate to the \gls{csi} only in the first round and it ignores the \gls{csi} observed in the subsequent rounds \cite[Sec.~V.D]{Sassioui17}. 

To remedy this problem and make \gls{harq} aware of the \gls{csi} in each round, various approaches have been proposed in the literature. One simple strategy suggested in \cite{Sassioui17} relies on the \gls{amc} rate-adaptation function, \ie $\R^\tnr{harq}(\SNR)=\R(\SNR)$ and applies the so-called \emph{packet dropping}, which in some cases terminates the \gls{harq} cycle before the $\kmax$-th round is reached and a \gls{nack} is received. 

This works as follows: in each round, the rate that might be offered by the \gls{amc}, $\R(\SNR_k)$ is compared to the rate chosen in the first round, $\R(\SNR_1)$. Rather than insisting on retransmitting the packet with a low nominal rate $\R(\SNR_1)$, \gls{phy} is allowed to take advantage of the observed \gls{snr} by transmitting a new packet with a high-rate $\R(\SNR_k)$. The packet is then dropped before the $k$-th round starts (and, instead, a first round of a new packet begins) \cite[Sec.~VI.A]{Sassioui17}
\begin{align}\label{drop.condition}
\R(\SNR_k)>\R(\SNR_1)\implies\text{Drop the packet}.
\end{align}

The packet-dropping is also ``backward compatible'' in the sense that it uses the same signaling as the conventional \gls{irharq} and occupies the same bandwidth. It eliminates the throughput penalty imposed by the \gls{irharq} but does not improve the throughput significantly. 

A natural question that arises is how the packet dropping \emph{combines} with a better rate-adaptation strategy $\R^{\tnr{harq}}(\SNR)$. Again, we would like to avoid the formal dependence on the channel model so we may use the simplified tuneable  adaptation defined in \eqref{delta.gamma} with $\Delta>1$. Indeed, by being more aggressive, we are exposed to a higher probability of undecoded packet, but we will mitigate this effect by  leveraging the very principle of retransmissions proper to \gls{harq}.

\begin{figure}[t]
\psfrag{SNR}[cc][cc][\siz]{$\SNRav$~[dB]}
\psfrag{throughput}[cc][cc][\siz]{Throughput $\bpcu$}
\psfrag{Erg}[lc][lc][\siz]{$\ov{C}$}
\psfrag{AMC}[lc][lc][\siz]{$\eta^{\amc}=\eta^{\IR}_1$}
\psfrag{HARQ}[lc][lc][\siz]{$\eta^{\IR}_{4}$}
\psfrag{HARQT}[lc][lc][\siz]{$\eta^{\IR}_{4}(\hat{\Delta})$}
\psfrag{DroppingX}[lc][lc][\siz]{$\eta^{\IR,\tr{drop}}_{4}(\hat{\Delta})$}
\psfrag{t=0.1XXXX}[lc][lc][\siz]{$\fD \tau=0.1$}
\psfrag{t=0.05}[lc][lc][\siz]{$\fD \tau=0.05$}
\begin{center}
\scalebox{\sizf}{\includegraphics[width=\linewidth]{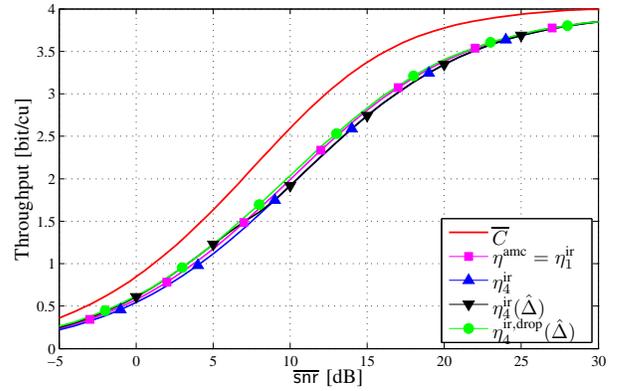}}
\end{center}
\caption{16-QAM over Rayleigh fading channels with  $\fD \tau=0.05$; the throughput of the conventional \gls{irharq} with the \gls{amc} rate adaptation, $\eta^{\IR}_{\kmax}$ as well as with the aggressive rate adaptation $\eta^{\IR}_{4}(\hat{\Delta})$ are compared to the throughput of \gls{irharq} with packet dropping $\eta^{\IR,\tr{drop}}_{4}(\hat{\Delta})$. The ergodic capacity, $\ov{C}$ \eqref{C.erg} is shown for reference.}\label{Fig:Rayleigh.conventional}
\end{figure}

\begin{example}[16-QAM, Rayleigh fading, threshold decoding]\label{Ex:QAM.Rayleigh}

To show an example of  throughput, we have to define how the decoding errors occur in \gls{harq}. To this end, we adopt the simplified threshold-decoding principle we used in \exref{ex:threshold.Doppler} and assume that the block length, $\Ns$, is sufficiently large so that the error occurs if the average accumulated \gls{mi} at the receiver is lower than the transmission rate \cite{Caire01}, that is, 
\begin{align}\label{err.k.def}
\Err_k=\SET{\sum_{l=1}^k I(\tilde{\SNRrv}_l)<\R(\SNRrv_1)}.
\end{align}

For $16$-\gls{qam}, the \gls{mi} function $I(\SNR)$ can be obtained numerically using the method from~\cite[Sec.~4.5]{Szczecinski_Alvarado_book}. Thus, the values of \gls{snr}s in \eqref{err.k.def} determine the decoding error $\Err_k$. 

We also consider here the idealized case of a continuous rates set, $\mcR=[0,4)$. Then, we obtain the \gls{amc}-optimal rate adaptation function $\R(\SNR)$ solving \eqref{eq:optimal/policy.K.1} with the \gls{per} function defined by  \eqref{per.Q1}, where i)~due to continuity of the rate set $\mcR$, we remove the index $[\ell]$)  
and ii)~we replace $\gamma_{[\ell]}^{\mr{th}}$ by $I^{-1}\big(\R(\SNR)\big)$.\footnote{We note here that, since we are considering only $16$-\gls{qam} modulation in this example, ``\gls{amc}'' has a somehow oxymoronic meaning of ``\gls{amc} with fixed modulation''. This is not a critical issue since we are interested in using the \gls{per} curves of the receiver, irrespectively how the coding/modulation is implemented.\label{footnote:AMC}}

\figref{Fig:Rayleigh.conventional} presents the throughput of the conventional \gls{irharq} compared to the ergodic capacity $\ov{C}$. The results indicate that increasing the number of rounds from $1$ to $4$ not only does not increase the throughput but, actually, is detrimental. Comparing with the similar results in \figref{Fig:Rayleigh.lharq0.95} this effect is less pronounced if we decrease the correlation between the estimated-  and the instantaneous \glspl{snr}, \ie  for a larger  normalized Doppler frequencies $\fD \tau$.

To explain these results, we can consider two successively observed \gls{snr}s, $\SNR_{1}$ and $\SNR_{2}$, such that $\SNR_{1}\ll \SNR_{2} \Longrightarrow \R(\SNR_{1})\ll R(\SNR_{2})$. We assume that a new \gls{harq} cycle starts with $\SNR_{1}$ and the first round experiences a decoding failure, \ie \gls{nack} is declared to the transmitter at the end of the first round. Using \gls{harq}, the same packet will be retransmitted with the maximum possible reward $\R(\SNR_{1})$. On the other hand, if only one transmission is allowed, \ie we use the \gls{amc}, the transmission of a new packet starts with a possible reward $\R(\SNR_{2})\gg \R(\SNR_{1})$. Thus, the reward of the \gls{amc} is larger than the one attainable by \gls{harq}. It happens because \gls{harq} ignores the \gls{csi} in the second transmission; this effect was analyzed in \cite{Sassioui17} where the penalty of using \gls{harq} in high \gls{snr} is formally demonstrated.

This also explains the rational behind the packet dropping. If $\R(\SNR_{2})>\R(\SNR_{1})$, according to \eqref{drop.condition}  we start the transmission of the new packet (that is, we drop/abandon the old one). The throughput of packet dropping, $\eta^{\IR,\tnr{drop}}_{4}(\hat{\Delta})$,  is also shown in \figref{Fig:Rayleigh.conventional}, where $\hat{\Delta}>1$ is the optimal ``aggressiveness'' factor, see \eqref{delta.gamma}, chosen to maximize the throughput for each value of the average \gls{snr} $\SNRav$, \ie $\hat{\Delta}$ is obtained as follow
\begin{align}
\hat{\Delta}=\argmax_{\Delta\in\Real^{+}}\eta\big(\R(\SNR\Delta)\big),
\end{align} 
where $\eta\big(\R(\SNR)\big)$ is the throughput using the rate adaptation policy $\R(\SNR)$. In the numerical examples, $\hat{\Delta}$ is obtained through the exhaustive research above the neutral value $\Delta=1$. \footnote{In practice, the exhaustive search is not feasible and the optimization is implemented via the on-line adaptation: $\Delta$ is fine-tuned using slow-feedback trying to maximize a parameter of interest---here the throughput. While $\Delta$ can take any positive value, the space of possible values $\Delta$ considered in the numerical example is logarithmically spaced between $1$ and $\Delta_{\max}$, where $\Delta_{\max}$ guarantees that $\R_{L}$ is always available for adaptation, \ie $\Delta_{\max}\cd \SNR \ge \gamma_{[L]}$ for all values of $\SNR$ we consider.}

As stated before, the packet dropping removes the detrimental effect of \gls{harq} but helps little in improving the throughput in its ``high'' range: for $\eta^\amc>2.5 \bpcuval$, we observe $\eta^\amc \approx \eta^{\IR,\tnr{drop}}_{4}(\hat{\Delta})$. 

We also show the throughput of the conventional \gls{harq} with the aggressive rate adaptation denoted by $\eta^{\IR}_{4}(\hat{\Delta})$, where no improvement over the conventional \gls{harq} $\eta^{\IR}_4$ is noted. Again, this is merely a confirmation of the formal statement made in \cite{Sassioui17}, which demonstrated that, for high \gls{snr}, irrespectively of the rate-adaptation policy, the throughput of \gls{harq} must deteriorate compared to the throughout of the \gls{amc}.
\end{example}

We see in \exref{Ex:QAM.Rayleigh} that being aggressive is helpful to \gls{harq} (and to \gls{harq} with packet dropping) in the low range of the throughput where we can use higher transmission rates  than those suggested by the \gls{amc} rate-adaptation $\R(\SNR)$. This provides another way of looking at the problem: the throughput cannot be improved for  high throughput values because there are no sufficiently high rates in the set $\mcR$; here $\max\set{\mcR}=4$ since a $16$-\gls{qam} modulation is used. However, extending the range of operation, that is, adding new rates to $\mcR$ will merely offset the problem: in any system, there is always a maximum transmission rate, and \gls{harq} will always fail to improve the throughput in the range that is close to $\max\set{\mcR}$.


\subsection{Upper Layers: Why We Ignore The Packet Loss at the PHY}\label{Sec:upper.layers}

\figref{fig:HARQ.AMC.model} shows two layers that are responsible for the retransmission: \gls{llc}, which implements the \gls{arq} protocol, and \gls{phy}, which implements \gls{amc} (and possibly also \gls{harq}). Since the \gls{phy} cannot guarantee a successful transmission, the packet contents is kept in the \gls{llc} buffer till a successful decoding is declared. In this way, some bits may require many \gls{phy} transmissions before being successfully decoded at the receiver. The introduction of \gls{arq} at \gls{llc} guarantees a zero-outage and yet does not affect the  throughput seen by the upper layers  \cite[Sec. II.A]{Jabi16}. Such a double-retransmission mechanism is adopted in~\gls{lte} \cite[Ch. 12]{Dahlman14_book}, where \gls{arq} is activated when dealing with loss-sensitive but delay-tolerant applications. 

Another scenario where the outage can be ignored is when the Fountain coding \cite{Ellis15} is used at the application layer: the $T$ application packets $\bb[t], t=1,\ld,T$  are transformed/encoded into a potentially unlimited number of information packets $\mfm[1], \mfm[2], \ld$. This is done in such a way that we can recover/decode all packets $\bb[t]$ from \emph{any} subset of $T+\tau$ packets $\mfm[n]$, where $\tau\ll T$. Thus, at the receiving end, we do not care if the decoding of \emph{each} of the packets $\mfm[n]$ succeeds but only require having $T+\tau$ of them decoded correctly. In other words, the throughput is what really counts and we can ignore the actual value of the packet loss probability.

\section{AMC and Layer-coded HARQ} \label{sec:lharq.principle}

In \exref{Ex:QAM.Rayleigh}, we explained how the retransmissions in \gls{harq} could penalize the throughput. We also identified the source of the problem: in the conventional \gls{harq}, the adaptation to the channel occurs only in the first transmission round when the \gls{amc}-like behavior is used to exploit the \gls{csi} $\SNR_1$; all subsequent rounds \emph{ignore} the \gls{csi}s  $\SNR_2,\ld, \SNR_\kmax$. 

On the other hand, the \gls{amc} alone discards the past channel observations $\by$ independently whether the decoding was successful or not. This clearly is a waste of resources and we want to combine now the adaptability of \gls{amc} and  the capacity of \gls{harq} to exploit the past observations.

\subsection{Variable-length \gls{harq} vs \gls{lharq}}\label{Sec:VLHARQ} 

The problem of throughput penalty occurring when \gls{harq} is combined with the \gls{amc} was partially remedied by the packet dropping, see \secref{Sec:HARQ}, where the \gls{csi} is used to decide whether to retransmit the undecoded packet using the entire channel block  or to drop it definitely. A \gls{vlharq}, studied in \eg \cite{Cheng03,Uhlemann03,Visotsky05,Pfletschinger10,Szczecinski13,Ramis11,Femenias09}, may be seen as a compromise between the very crude adaptation of packet dropping and non-adaptive, conventional \gls{harq}. The idea is to encode an information packet $\mfm$ into a sequence of subcodwords $\bx_1,\ld, \bx_\kmax$,  whose length, \ie the number of symbols they are composed of, $\Nsk{k}$, may vary from one subcodeword to another. We assume that the number of symbols in the first codeword is fixed $\Ns=\Nsk{1}$, \ie $\bx_1\in \Real^{\R_1\Ns}$, and for convenience, we normalize the length of the remaining ones via $\ell_{\kk}\triangleq N_{\tr{s},\kk}/\Ns$. 

The optimization problem we have to solve is to find the values $\R_1,\ell_{2},\ld,\ell_{\kmax}$ which maximize the throughput of \gls{vlharq}. However, we cannot simply adopt the solutions developed previously in \cite{Pfletschinger10,Szczecinski13} because they did not deal with the instantaneous \gls{csi}; others, \eg \cite{Ramis11,Femenias09}, targeted particular values of the \gls{per}, which is not of concern in our paper. More importantly, in this work we admittedly ignore the knowledge of the distribution of the \gls{snr}; see comments after \eqref{adapt.func}. 

Therefore, for comparison purpose we will propose a new approach which is detailed in \exref{Ex:VL.vs.L}. What is important to understand at this point is that any adaptation will be based on the \gls{per} function which is defined for each \gls{harq} round
\begin{align}\label{PER.function.MD}
\PER_k&(\SNR_k, \ell_k;\cd)\nonumber\\
&\triangleq \PER\left(\SNR_k, \ell_k;\set{\tilde{\SNR}_t}_{t=2}^{k-1},\R_{1},\set{\ell_{t}}_{t=2}^{k-1} \right)
\end{align}
and which depends on the following arguments:
\begin{itemize}
\item the \gls{snr} estimated before the $k$th transmission round, $\SNR_k$; 
\item the length, $\ell_k$ of the subcodeword $\bx_k$ used in the the $k$th round;
\item all past channel \gls{snr}s, $\tilde{\SNR}_1,\ld,\tilde{\SNR}_{k-1}$; and 
\item all transmission parameters chosen in the previous rounds, \ie $\R_{1}(\SNR_1)$, and $\ell_{t}(\SNR_{t}), t=2,\ld, k-1$. 
\end{itemize}

It is immediately clear that, in the case of the practical encoders/decoders, finding empirically the \gls{per} function \eqref{PER.function.MD} with so many parameters presents a considerable challenge. This is the main reason why the previous works on the variable-length \gls{harq} used simplifying assumptions to take care of its  large number of arguments. For example, \cite{Szczecinski13} assumed that the receiver's \gls{per} function is binary and the error events are defined in a way similar to \eqref{err.k.def};\cite{Femenias09,Sassioui17} used analytical \gls{per} functions which have to be fit to the actual decoding curves. Similarly, while the empirical decoding curves are used  in \cite{Visotsky05,Pfletschinger10} the approximations are still necessary to predict the performance for arbitrary values of the \gls{snr} and the codeword length.

Besides the practical difficulty of describing the \gls{per} function, there is also another important, system-level consideration: \gls{vlharq} uses  only fraction of channel resources which goes against the goal of assigning fixed resources to each of the \gls{harq} rounds. Additional assumptions are then required. In particular,  \cite{Liu04,Femenias09} let the user to transmit variable number of  packets within a fixed-length \emph{frame}. This, however, raises a problem of resource assignment  to many packets in the frame as well as implies the signalling overhead: the length of the codeword of each packet in the frame must be indicated.

\gls{lharq}, addresses the above difficulties as follows: instead of transmitting the new (redundant) coded symbols $\bx_k, k=2,\ld, \kmax$ to enable the decoding of the packet $\mfm_1$, transmitted in the first round, we rather transmit (some of) the bits of the packet $\mfm_1$ (we denote these redundant bits as $\mfm_1'$), and we ``fill'' the remaining space with a fresh content $\mfm_2$. So the conventional channel coding is preceded by the mixing of the punctured information packets; this is the idea of layered-coding, \gls{lharq}, proposed in \cite{Popovski14} and modified in \cite{Jabi17b}. The decoding is also done in layers, where the  mixed packets $\mfm'_1$ and $\mfm_2$ are decoded first and, next, the packet $\mfm_1$ is recovered with the aid of $\mfm'_1$.


The main advantages are that {\it{i)}}~the channel coding is not affected by the packet mixing and can be implemented without joint-coding considerations, {\it{ii)}}~the decoding is straightforward and can be done with commercially available decoders as demonstrated in \cite{Jabi17b}, and finally {\it{iii)}}~the decoding results are described  by one-dimensional  \gls{per} curves which depend only on one \gls{snr}; the dependence of the decoding results on other \gls{snr}s is implicit and defined via recursive (layerer) decoding.

Using fixed resources at the \gls{phy} may be seen as a way to reduce the signaling overhead and to ease the resource management (required, \eg for the scheduling in multi-user systems \cite[Ch.~12.1]{Dahlman14_book}). A cautionary note is in order here: the choice of a particular transmission strategy is a complex issue which depends on many system-level elements that are difficult to  include in the \gls{phy}-level analysis/design.

\subsection{\gls{lharq} Example: $\kmax=2$}\label{sec:sbrq.principle}
To explain the details, it is easier to start with the simplest case of \gls{harq}, \ie when $K=2$. 

\textbf{Transmitter}

The first transmission is done as in the conventional \gls{amc}: using the rate $\R_{1}$, the packet $\mfm_1 \in\set{0,1}^{R_1 \cd \Ns}$ is encoded and the resulting codeword, $\bx_1=\Phi[\mfm_1]$ is transmitted over  the channel. Then, if the decoding succeeds, $\hat{\mfm}_1=\mfm_1$ (an \gls{ack} is fed back), the earned reward is given by $\mfR[n]=\R_1$ and we move to the next packet. If the decoding fails, $\mfR[n]=0$. The average reward attainable in the first round is thus the same as in the case of the \gls{amc}: $\mfR[n]=\R_1\big(1-\PER(\SNR_1;\R_1) \big)$.


When the decoding fails, \ie $\Err_1$ is observed, in the next channel block, we implement the second round of \gls{harq}. First, we choose the channel coding rate $\R_{2}$. This is similar to the conventional \gls{amc}. However, we now encode the packet  $\mfm_{[2]}$  which is a mix of two subpackets: $\mfm'_{[1]}$,  which  contains ``old'' bits taken from the unsuccessfully transmitted packet $\mfm_1$,\footnote{In other words, $\mfm'_{[1]}$ is a punctured version of $\mfm_1$.} and the subpacket $\mfm_2$, composed of ``new'' bits
\begin{align}\label{mfm2.2}
\mfm_{[2]}&=[\mfm'_{[1]},\mfm_2]\in\Binary^{\R_2\Ns}\\
\mfm'_{[1]}&=\Phi_1^\tr{b}[\mfm_1]\in\Binary^{\Rs{1}\Ns},
\end{align}
where $\Rs{1}$ is the packet-mixing rate, \ie $\mfm'_{[1]}$ contains $\Rs{1}\Ns$ bits of $\mfm_{1}$, and therefore, the natural constraint here is $\Rs{1}<\min\set{\R_1,\R_2}$: the packet $\mfm'_{[1]}$ cannot contain more bits that $\mfm_{[2]}$  or  $\mfm_1$.

The channel coding is done as before, \ie $\bx_2=\Phi[\mfm_{[2]}]$.

\textbf{Receiver}

At the receiver, initially, we ignore the particular structure of the packet $\mfm_{[2]}$, and we decode it as $\hat{\mfm}_{[2]}=\tr{DEC}[\by_2]$. Again, we obtain the reward $\mfR[n]=\R_2$ or $\mfR[n]=0$; this is how the conventional \gls{amc} would work. However, if we decode the packet $\mfm_{[2]}$, we may exploit the structure \eqref{mfm2.2}. In particular, $\mfm'_{[1]}$ provides a priori information about $\mfm_1$ and can be used in the ``backtrack'' decoding
\begin{align}\label{dec.back}
\hat{\mfm}^\tnr{b}_{[1]}=\tr{DEC}[\by_1;\mfm'_{[1]}]
\end{align}
as proposed in \cite{Popovski14}. We define the backtrack decoding error as \mbox{$\Errb{1}\triangleq\set{\hat{\mfm}^\tnr{b}_{[1]}\neq \mfm_{[1]}}$} and describe its probability by the backtrack \gls{per} function
\begin{align}\label{PER.back}
\PER^\tnr{b}(\tilde{\SNR};\R,\Rs{})
&\triangleq\PR{ \Errb{}|\Err, \tilde{\SNRrv}=\tilde{\SNR},\R,\Rs{}}\\
\label{PER.back.2}
&=\frac{\PR{\Err\wedge \Errb{}|\tilde{\SNRrv}=\tilde{\SNR},\R,\Rs{}}}
{\PR{\Err|\tilde{\SNRrv}=\tilde{\SNR},\R}},
\end{align}
which depends  on the \gls{snr}, $\tilde{\SNR}$, that was actually observed in the first round.

The \gls{per} function \eqref{PER.back} must decrease with $\Rs{}$: from the information-theoretic point of view, the decoding \eqref{dec.back} can  succeed, even if the decoding $\tr{DEC}[\by_1]$ has failed, because the uncertainty about the packet $\mfm_1$ (\ie the rate at which the unknown bits are coded) is decreased. From the practitioner's point of view, knowing $\mfm'_{[1]}$, we have absolutely certain a priori information about the information/systematic bits $\mfm'_{[1]}$ from the packet $\mfm_1$, which improves the performance of the decoder.\footnote{For example, in iterative binary decoders, knowing some of systematic bits corresponds to setting the corresponding \glspl{llr} to $\pm\infty$.} 

Since we may now  recover the entire packet $\mfm_1$ (having $\mfm'_{[1]}$ means that $\mfm_1$ was partially decoded in  the second round), the reward for the second cycle is calculated as
\begin{align}
\nonumber
\mfR[n]=&\Big[\R_2  +  (\R_1-\Rs{1})\big(1-\PER^\tnr{b}(\tilde{\SNR}_1;\R_1,\Rs{1})\big) \Big]\\
\label{reward.k=2}
&\quad\big(1-\PER(\SNR_2;\R_2)\big),
\end{align}
here, since $\mfm'_{[1]}$ is common in both transmission rounds,  the associated reward, $\Rs{1}$,  is already included in $\R_2$; thus, we subtract $\Rs{1}$ from $\R_1$ to avoid counting this reward twice.

The only remaining question is how to choose the rates $\R_1$, $\R_2$, and $\rho_1$, but we postpose the discussion untill we formulate the solution for an arbitrary number of rounds.

\subsection{\gls{lharq}: General Case}\label{sec:sbrq.g}
 
\begin{figure}[tb]
\hspace*{-0.75cm} \scalebox{0.63}{

\usetikzlibrary{shapes,arrows}

\tikzstyle{decision} =  [rectangle, draw, fill=gray!15, 
    text width=12em, text centered, rounded corners=15pt, minimum height=4em]
 \tikzstyle{decision1} = [diamond, draw, fill=blue!20, 
  text width=12.5em, text badly centered, node distance=3cm, inner sep=0pt]
\tikzstyle{decision1} = [rectangle, draw, fill=gray!20, 
    text width=15em, text centered, rounded corners=25pt, minimum height=4em]
    \tikzstyle{block} = [rectangle, draw, fill=gray!20, 
    text width=7em, text centered,  minimum height=4em]
\tikzstyle{block1} = [rectangle, draw, fill=gray!20, 
    text width=22em, text centered,  minimum height=4em]
\tikzstyle{block2} = [rectangle, draw, fill=gray!20, 
    text width=15em, text centered,  minimum height=1em]
     \tikzstyle{block3} = [rectangle, draw, fill=gray!20, 
    text width=11em, text centered,  minimum height=0.5em]
     \tikzstyle{block3fork} = [diamond, draw, fill=gray!20, text width=4em, 
    text centered,  minimum height=0.5em]
       \tikzstyle{block4} = [rectangle, draw, fill=gray!20, 
    text width=14em, text centered,  minimum height=0.5em]
    \tikzstyle{block5} = [rectangle, draw, fill=gray!20, 
    text width=14em, text centered,  minimum height=0.5em]
    
\tikzstyle{line} = [draw, -latex']
\tikzstyle{cloud} = [draw, ellipse,fill=red!20, node distance=3cm,
    minimum height=2em]
    \tikzstyle{arrow} = [thick,->,>=stealth]
     \tikzstyle{noarrow} = [thick,-,=>stealth]
      \tikzstyle{form_back_t} = [draw, minimum width=9cm, rounded corners, text width=0.8cm,fill=red!15, 
text centered,  minimum height=1cm, inner ysep=15pt]
  \tikzstyle{form_back_c} = [draw, minimum width=3.5cm, rounded corners, text width=3.5cm,fill=blue!15, 
text centered,  minimum height=1.5cm]
 \tikzstyle{form_back_r} = [draw, minimum width=11.85cm, text width=0.8cm,rounded corners, fill=green!15, 
text centered,  minimum height=1cm,inner ysep=15pt]

\begin{tikzpicture}[node distance = 2cm, auto]
    

        \node [block1] (init) {\textbf{Packet mixing}:
\begin{Sequation}    
   \begin{align}
\label{mfm.p}
\mfm'_{[k-1]}&=\Phi^\tnr{b}_{k-1}[\mfm_{[k-1]}]\in\set{0,1}^{\Rs{k-1}\Ns} \\
\mfm_{[k]}&=[\mfm'_{[k-1]},\mfm_k]\in\set{0,1}^{\R_{k}\Ns} \label{mfm.k}
\end{align}
\end{Sequation}
  };
    
\node [block4, below of=init, xshift=0cm, yshift=-0.5cm] (init1) {\textbf{Channel coding}
\begin{Sequation}
\begin{align}\label{bx.k}
\bx_k&=\Phi[\mfm_{[k]}]
\end{align}
\end{Sequation}
    };
      \node [block3, above of=init, xshift=2.25cm, yshift=0.5cm] (new) {\textbf{New \gls{harq} cycle:}
      \begin{Sequation}
       \begin{align} \label{eq:new.cycle.G}
       k=1; \Rs{0}=0
       \end{align}
       \end{Sequation}
       };
          \node [block3, above of=init, xshift=-2.25cm, yshift=0.5cm] (retran) {\textbf{New \gls{harq} round:}
          \begin{Sequation}
       \begin{align}
       \label{eq:backtrack.rate.G}
       k=k+1
       \end{align}
       \end{Sequation}
       };

   \draw [arrow] (retran) -- (init);
   \draw [arrow] (new) -- (init);
   \draw [arrow] (init) -- (init1);

       
        \node [form_back_c, below of=init1, xshift=0cm, yshift=-0.5cm] (chan) {\bf{Channel}
   \begin{align}
\by_k=\sqrt{\tilde{\SNR}_k} \bx_k+\bz_k\nonumber
\end{align}
        };

           \node [block4, below of=chan, xshift=2.75cm, yshift=-0.85cm] (Dec) {\textbf{Decoding}
\begin{Sequation}
       \begin{align}\label{dec.mk.algo}
      \hat{\mfm}_{[k]}=\tnr{DEC}[\by_k]
       \end{align}
   \end{Sequation}
       };

    \node [block3fork, below of=Dec, xshift=0cm, yshift=-0.1cm] (Err) { $\Err_k$ };
               
       \node [block4, below of=Err, xshift=0cm, yshift=-1.7cm] (backtrack) {\textbf{Backtrack decoding}
\begin{Sequation}
   \begin{align}
   &\text{for}~l=k-1, \ld, 1\nonumber\\
   \label{Backtrack.decoding}
&~\hat{\mfm}^\tr{b}_{[l]}=\tnr{DEC}[\by_{l}, \mfm'_{[l]}],\\
\label{stop.backtrack}
&~ \Err_l^\tr{b} \implies\text{stop}
\end{align}
\end{Sequation}
       };

       \node [block4, left of=Err, xshift=-3.25cm] (next) {\textbf{Mixing rate adapt. (HARQ)}
\begin{Sequation}
       \begin{align}\label{eq:retans.cond.G}
       \Rs{k}=\Rs{k}(\R_k,\tilde{\SNR}_k)
       \end{align}
\end{Sequation}
       };
       
        \node [block3fork, below of=next, xshift=0cm, yshift=-0.1cm] (Check_rho) {   \mbox{$\Rs{k} > \R_{k+1}$}   };
        
        \node [block4, fill=gray!50, below of=Check_rho, xshift=0cm, yshift=-0.1cm] (rate_Adp) {\textbf{Rate adapt. (AMC)}
\begin{Sequation}
    \begin{align}
\R_{k+1}&=\R(\SNR_{k+1})
\label{R.k1}
\end{align}
\end{Sequation}
       };

      \begin{pgfonlayer}{background}       
      \node[ form_back_r, fit=(rate_Adp)(Dec)(backtrack)(next)(Check_rho)]   (Recei) {}; 
      \node[below] at (Recei.north) {\bf{Receiver}};  %
     
      \node[ form_back_t, fit=(init)(init1)(new)(retran)  ]   (Trans) {}; 
      \node[below] at (Trans.north) {\bf{Transmitter}};  %
       \end{pgfonlayer}

       \path (Check_rho.west)+(-0.3,0.2)  node {no}; 
       \path (Check_rho.east)+(0.3,0.2)  node {yes}; 

       \draw [arrow] (Trans) -- (chan);
        \draw [arrow] (chan) -- (Recei);
       \draw [arrow] (Dec) -- (Err);

\draw [arrow] (Err.west) -- node[anchor=north west] {yes} (next);
 \draw [arrow] (Err.south) -- node[anchor=south east] {no} (backtrack);

       \draw [arrow] (next) -- (Check_rho);
       \draw [arrow] (rate_Adp) -- (Check_rho);

       \path (backtrack.east)+(0.5,0)  node(P1) {}; 
       \draw [noarrow] (P1.base) -- (backtrack.east);
       \draw [arrow] (P1.base) |- (new.east); 
       \draw [noarrow,dashed] (Check_rho.east)-| (P1.base); 

        \path[noarrow] (Check_rho.west)+(-1.6,0) node (P2) {}; 
        \draw[arrow] (P2.base) |- (retran.west);
        \draw[noarrow] (P2.base) -- node{}(Check_rho.west);

\end{tikzpicture}}
\caption{Actions flow at the transmitter and the receiver in the proposed \gls{lharq}. The rate adaptation at the receiver is composed of two parts: the mixing rate adaptation \eqref{eq:retans.cond.G} proper to \gls{harq} which follows the decoding error $\Err_k$, and the channel rate adaptation \seqref{R.k1} proper to the \gls{amc}. The adaptation results are used by the transmitter independently in two encoding steps \seqref{mfm.p} and \seqref{mfm.k}. Their dependence shows up only if $\Rs{k} >\R_{k+1}$, which may terminate the \gls{harq} cycle, as shown by the dashed transition line at the receiver.}\label{Fig:L.HARQ.G}
\end{figure}
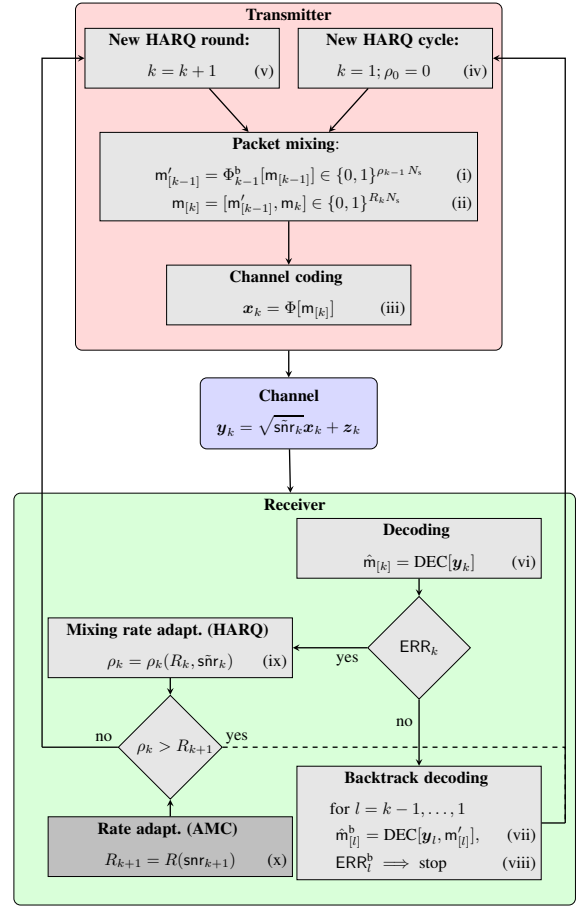

We now generalize the encoding/decoding to $\kmax>2$; we explain and separate the encoding/decoding and adaptation operations in the flowchart shown in  \figref{Fig:L.HARQ.G}.

\textbf{Transmitter}

In each round, the encoding is done in two steps. First, we do a binary packet mixing: the old packet $\mfm_{[k-1]}$ is punctured, using rate $\Rs{k-1}$, yielding a subpacket $\mfm'_{[k-1]}=\Phi^\tnr{b}_{k-1}[\mfm_{[k-1]}]$ according to \seqref{mfm.p} in \figref{Fig:L.HARQ.G};\footnote{The equations on \figref{Fig:L.HARQ.G} are labeled with a roman numerology and this is the reason we explicitly use the abbreviation ``Eq.".} the resulting $\mfm'_{[k-1]}$ is concatenated with new information bits $\mfm_k$ to form the packet $\mfm_{[k]}=[\mfm'_{[k-1]},\mfm_k]$, see \seqref{mfm.k}. Second, the channel encoding: \seqref{bx.k} straightforwardly encodes $\mfm_{[k]}$ using the rate $\R_{k}$. 

\textbf{Receiver}

Using only the observation $\by_k$, the receiver starts the first step of the decoding process  \seqref{dec.mk.algo}. 

\begin{itemize}

\item If the packet $\mfm_{[k]}$ is decoded correctly in the $k$-th round, \ie $\hat{\mfm}_{[k]}=\mfm_{[k]}$, we recover the message $\mfm'_{[k-1]}$ and $\mfm_k$ according to \seqref{mfm.k}. With $\mfm'_{[k-1]}$ and $\by_{k-1}$ known in the $k$-th round, we ``backtrack'' decode the packet $\mfm_{[k-1]}$ according to~\seqref{Backtrack.decoding}. If the backtrack decoding $\mfm_{[k-1]}$ is successful, \ie $\hat{\mfm}_{[k-1]}^\tnr{b}=\mfm_{[k-1]}$, we then recover ${\mfm}_{k-1}$ and $\mfm_{[k-2]}'$ according to~ \seqref{mfm.k}. In this way, we continue the backtrack decoding to recover all the packets $\mfm_{k-2},\ld, \mfm_1$. However, if a backtrack decoding error, $\Err^\tnr{b}_l=\set{\hat{\mfm}^\tnr{b}_{[l]}\neq\mfm_{[l]}}$, is observed for any packet $l\in\set{k-1,\ld,1}$, we abandon the decoding and declare the packets from the previous rounds, $\mfm_{l}, \mfm_{l-1},\ld, \mfm_{1}$ lost.

\item If an error 
\begin{align}\label{}
\Err_k=\set{\hat{\mfm}_{[k]}\neq\mfm_{[k]}}
\end{align}
is observed, we have to prepare the packet $\mfm_{[k+1]}$ for the next \gls{harq} round. To this purpose and according to \seqref{mfm.p} and \seqref{mfm.k}, we need to determine the mixing $\Rs{k}$ and the transmission rate $\R_{k+1}$, respectively. In particular, we consider that $\Rs{k}$ is a function of $\R_{k}$ and $\tilde{\SNR}_k$, \ie $\Rs{k}\equiv\Rs{k}(\R_{k},\tilde{\SNR}_k)$ and the way to obtain the adaptation function $\Rs{k}(\R_{k},\tilde{\SNR}_k)$ will be explained in \secref{Sec:simple.rate.adaptation}. On the other hand, the transmission rate $\R_{k+1}$ is available only after the receiver observes/estimates $\SNR_{k+1}$, see \seqref{R.k1}. Finally, we compare $\Rs{k}$ to the rate of the next transmission $\R_{k+1}$ in order to decide whether the new \gls{harq} round can be done, see \seqref{eq:backtrack.rate.G}, or the \gls{harq} cycle should be restarted, see \seqref{eq:new.cycle.G}.

\end{itemize}

We emphasize here that the two-steps decoding adopted in \gls{lharq} is not the only option: it is possible for instance, to perform the joint decoding of all packets ${\mfm}_{1}, \ld, {\mfm}_{k}$ using all channel outcomes $\by_1,\ld,\by_k$ as proposed in \cite{Jabi17a}.\footnote{Another decoding scheme was also proposed in \cite[footnote 2]{Jabi17b}.} However, motivated by the fact that the simplified encoding-decoding schemes adopted in \gls{lharq} do not impose any throughput penalty \cite[Th. 3]{Trillingsgaard17}, and  to keep the backtrack decoding simple, we do not consider any of those possibilities here.

\subsection{Optimal Rate Adaptation}\label{Sec:Rate.adaptation}

While the operations of the transmitter and the receiver are simple, we still must adapt the transmission rates $\R_k$ and the mixing rates $\Rs{k}$. For the purpose of the discussion, it is convenient to write clearly the rewards earned in each round, which generalizes \eqref{reward.k=2}. Namely,
\begin{align}
\label{reward.Rk}
\mfR_k&=(\R_k  + \mfJ_{k}  )\big(1-\PER(\SNR_k;\R_k)\big)\\
\nonumber
\mfJ_{k}&=(\mfJ_{k-1}+\R_{k-1}-\Rs{k-1})\\
\label{reward.Jk}
&\qquad ~\qquad\big(1-\PER^\tnr{b}(\tilde{\SNR}_{k-1};\R_{k-1},\Rs{k-1})\big),
\end{align}
where $\mfJ_k$ is the reward that can be obtained in the $k$-th \gls{harq} round thanks to the backtrack decoding. 

The above allows us to formulate the problem using a \gls{mdp} formalism, where we define the \emph{states} as a tuplet
\begin{align}\label{s.k.def}
\mfs_k\triangleq (\mfJ_{k-1}, \SNR_k, \tilde{\SNR}_{k-1}, \R_{k-1}),
\end{align}
and the objective is to find the optimal \emph{actions} for each state $\mfs_k$
\begin{align}\label{}
\mfa_k\triangleq(\R_k,\Rs{k-1})=\pi (\mfs_k),
\end{align}
where $\pi(\cd)$ is the adaptation function or -- the \emph{policy}. Since knowing the action $\mfa_k$ determines the transition probability from state $\mfs_k$ to state $\mfs_{k+1}$, and the reward depends solely on the state and the action, the optimal policy which maximizes the reward can be found using known algorithms, such as policy iteration or value iterations \cite[Chap. 7]{Bertsekas05_book}.

While, the \gls{mdp} leads to a very efficient optimization, its implementation  may still be challenging if the dimension of the space is large. Here, unfortunately, this is the case: even if we assume that the available transmission rates, $\R_k$, are discrete, the three arguments remaining  in \eqref{s.k.def} are continuous. The problem is thus hardly tractable numerically. 

Moreover, we can argue that, while finding the optimal rate adaptation functions is theoretically interesting, its practical value is much less important because {\it{i)}}~the optimal solution would depend on the fading distribution, \eg defined by \eqref{pdf.SNR}, which we consider impractical since the latter is not known, see comments after \eqref{eq:optimal/policy.K.1};  and {\it{ii)}}~the rate adaptation function would be multidimensional  (would depend on as many parameters as there are dimensions in the state of the \gls{mdp}; here, four dimensions), which is not only tedious to implement but also makes the \gls{amc} and \gls{harq} adaptations coupled; which goes against the simplicity of the \gls{amc} adaptation which we consider a valuable feature.

Therefore, the simplifications and the heuristic solutions we propose may turn out to be not only viable but also more \emph{practical} alternatives to the complex  solutions that can be obtained from the optimization defined by the \gls{mdp} framework. Having said this, we believe it would be interesting to find these optimal solutions; first, to verify how far our heuristic solution falls from the optimality; second, to obtain a better insight into the importance of the observed parameters of the state $\mfs_k$ and, possibly, to derive other useful heuristics.

\subsection{Simplified Rate Adaptation}\label{Sec:simple.rate.adaptation}

We will propose simple, and intuitively justified heuristics, where we target the independence of adaptation strategies between \gls{amc} (here, the choice of $\R_k$)  and \gls{harq} (here, the choice of the packet mixing rate $\Rs{k-1}$). 

First of all, we decide to adapt the coding rates in the same way irrespectively of the past \gls{harq} transmission rounds
\begin{align}\label{amc.2.rounds}
\R_k(\SNR)=\R(\SNR),
\end{align}
where $\R(\SNR)$ is the \gls{amc} rate-adaptation policy we found in \eqref{eq:optimal/policy.K.1} and which we can also eventually enhance with the fine-tuning \eqref{delta.gamma}. This can be seen as a maximization of the instantaneous reward $\mfR_k$ in \eqref{reward.Rk} setting $\mfJ_k=0$.

Second, we have  to decide what packet mixing rate $\Rs{k-1}$ should be used in the $k$-th round. The first simplification due to the \gls{amc} approach \eqref{amc.2.rounds} is that we make $\Rs{k-1}$ independent of $\SNR_k$, \ie 
\begin{align}\label{}
\Rs{k}=\Rs{k}(\R_{k},\tilde{\SNR}_{k}). 
\end{align}

Next, from \eqref{reward.Jk},  we see that small $\Rs{k-1}$ increases the potential reward (given by $\R_{k-1}-\Rs{k-1}$ in \eqref{reward.Jk}), yet it has to be sufficiently large to make the backtrack \gls{per}, $\PER^\tnr{b}(\tilde{\SNR}_{k-1};\R_{k-1},\Rs{k-1})$, low. The latter requiremement is also needed to recover all the packets from the previous rounds because we decided to abandon all of them once we observe the backtrack decoding error, $\Err^\tnr{b}$, see \seqref{stop.backtrack}.

To strike a balance between these two requirements, we decide to choose the minimum packet-mixing rate to attain the prescribed value of the backtrack \gls{per}, $\epsilon_\tnr{b}$
\begin{align}\label{rho.1.heuristic}
\Rs{k-1}=
\begin{cases}
0 \qquad ~\qquad ~\qquad \text{if}\quad \PER^\tnr{b}(\tilde{\SNR}_{k-1};\R_{k-1},\R_{k})>\epsilon_\tnr{b}\\
\underset{\Rs{}<\R_{k}}{\min}\set{ \Rs{}\in \mcR_\tnr{b}: \PER^\tnr{b}(\tilde{\SNR}_{k-1};\R_{k-1},\Rs{})\leq\epsilon_\tnr{b}}\quad\text{o.w.}
\end{cases}
\end{align}
where the first condition line in \eqref{rho.1.heuristic} takes care of the case when it is impossible to attain the \gls{per} value of $\epsilon_\tnr{b}$;\footnote{The \gls{per} function $\PER^\tnr{b}(\tilde{\SNR}_{k-1};\R_{k-1},\Rs{})$ is decreasing with $\Rs{}$, and attains its minimum for the maximum admissible value of $\Rs{}=\R_{k}$.} then, we set $\Rs{1}=0$, which means that $\mfm_{[k]}=\mfm_k$, see \seqref{mfm.p}--\seqref{mfm.k}, and this is equivalent to dropping the packet $\mfm_{[k-1]}$ and starting a new \gls{harq} cycle; $\mcR_\tnr{b}$ denotes the set of available mixing rates. 

Since, \emph{after} the $k$-th round, the receiver knows both the value of $\tilde{\SNR}_{k}$ and $\R_k=\R(\SNR_k)$, it can calculate also $\Rs{k}$ via \eqref{rho.1.heuristic}, and, before the round $k+1$ starts, $\Rs{k}$ can be sent to the transmitter together with $\R_{k+1}$ using the feedback $\mcF$.

The structure of the transmitter and the receiver is shown in \figref{Fig:L.HARQ}. The important element is that the coding and the adaptation to the \gls{csi}/\gls{snr} is done by the \gls{amc} independently of the operation of \gls{harq}. We  also note that we still maintain a certain degree of adaptability with a fine-tuning \eqref{delta.gamma} for the rate-adaptation in the \gls{amc} controller, and with the possible adjustment of the target backtrack \gls{per}, $\epsilon_\tnr{b}$.

\begin{figure}[bt]
\begin{center}
\scalebox{0.63}{\input{figures/MM_model_V8}}
\end{center}
\caption{Encoding and decoding in \gls{lharq}. The \gls{amc} controller adjusts the rate of the channel encoder, $\Phi_k$, independently of the \gls{harq} which selects the packet-mixing rate (\ie the rate of the puncturer $\Phi^\tnr{b}_k[\cd]$) . The decoding is done block-by-block using the channel outcome $\by_k$ and $\hat{\mfm}'_{[k]}$, \ie $\hat{\mfm}^\tr{b}_{[k]}=\tnr{DEC}[\by_k,\hat{\mfm}'_{[k]}]$. We assume here that $\kmax=3$ rounds are carried out, where $\Err_1$ and $\Err_2$ are observed (the first and the second rounds are in error) and the third \gls{harq} round is successful, \ie $\hat{\mfm}_{[3]}=\tnr{DEC}[\by_3]=\mfm_{[3]}$; the backtrack decoding is then successful as well, which means that $\hat{\mfm}^\tr{b}_{[k]}=\tnr{DEC}[\by_k; \hat{\mfm}'_{[k]}]={\mfm}_{[k]}, k\in\set{2, 1}$.}\label{Fig:L.HARQ}
\end{figure}

\section{Examples}\label{Sec:examples}

 We note first that the operation of the \gls{lharq} is solely based on the conventional \gls{amc} and the added step of packet mixing. We emphasize again that the formulas presented up to now do not take into account any prior knowledge about the statistical behavior of the channel. That is, while we will use Rayleigh fading \eqref{pdf.SNR} to model the \gls{snr}s, we do not exploit this knowledge in the design of \gls{harq}.\footnote{Although   the joint distribution \eqref{eq:cond.probablity} is used in \eqref{PER.func.Ex} to define the \gls{per} function, this is done for the sake of numerical comparison. In practice we will rather use the thresholds $\gamma_{[\ell]}$ estimated by the transmitter as in \eqref{gamma.ell.eps}; this can be done via slow-feedback adaptation measuring the value of the \gls{per} without knowing the entire \gls{per} function.}

\begin{figure}[tb]
\psfrag{SNR}[cc][cc][\siz]{$\SNR_k$~[dB]}
\psfrag{throughput}[cc][cc][\siz]{$\R_k=\R(\SNR \Delta)$}
\psfrag{ AMCC}[lc][lc][\siz]{$\Delta=1$}
\psfrag{ LHAQR2}[lc][lc][\siz]{$\Delta=\hat{\Delta}, \kmax=2$}
\psfrag{ LHARQ4XXX}[lc][lc][\siz]{$\Delta=\hat{\Delta},\kmax=4$}
\begin{center}
\scalebox{\sizf}{\includegraphics[width=\linewidth]{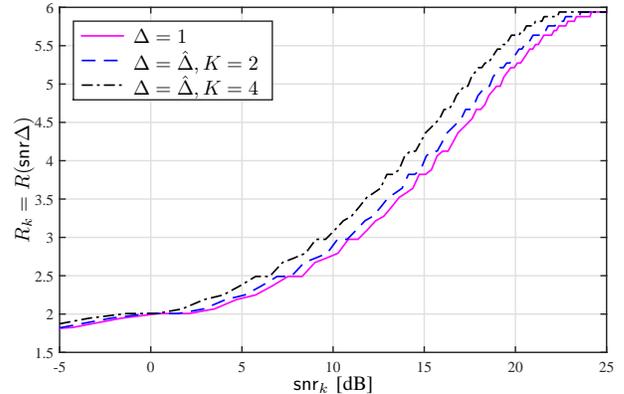}}
\end{center}
\caption{ Example of the optimal rate policy of \gls{lharq} when $\fD \tau=0.05$ and $\SNRav=20$ dB. $\Delta=0$ means that the \gls{amc} policy \eqref{adapt.func} is used; $\Delta=\hat{\Delta}$ means aggressive adaptation \eqref{delta.gamma} with an optimized aggressiveness factor.}\label{Fig:Rayleigh.policy.lharq0.95}
\end{figure}

\begin{example}[16-QAM \& 64-QAM, Rayleigh fading, threshold decoding, continuation of \exref{Ex:QAM.Rayleigh}]\label{ex:lharq.rate.policy}

Using the threshold decoding approach from \exref{Ex:QAM.Rayleigh}, the decoding errors are deterministically defined by the \gls{mi} in \eqref{err.k.def} and interpreting the knowledge of $\mfm_{[1]}'$ as decrease of the information rate from $\R_1$ to $\R_1-\Rs{1}$, the error event is defined as 
\begin{align}\label{errk.mi}
\Errb{k}= \set{ I(\tilde{\SNR}_k)<\R_k-\Rs{k} }.
\end{align}

Consequently, we can transform \eqref{rho.1.heuristic} into 
\begin{align}\label{rho.k.MI}
\Rs{k}&=\R_k-I(\tilde{\SNR}_{k}),
\end{align}
which holds for any $\epsilon_\tnr{b}\geq 0$ if we assume that the set of backtrack rates, denoted $\mcR_\tnr{b}$, is continuous, \ie $\mfR_\tnr{b}\in[0,\R_k)$. This idealized assumption will be relaxed in \exref{Ex:TC}.

With the \gls{amc}-optimal rate-adaptation policy $\R(\SNR)$, having the rate-mixing function \eqref{rho.k.MI}, as well as the decoding error events defined by \eqref{err.k.def} and \eqref{errk.mi}, we can run the Monte-Carlo simulation. The throughput of the proposed \gls{lharq}, $\eta^{\Lc}_{\kmax}$ is shown in \figref{Fig:Rayleigh.lharq0.95},  where we compare it  with \gls{irharq} for two different values of the normalized Doppler frequency, $\fD \tau\in\set{0.1,0.05}$. We assume that both $16$ and $64$-\gls{qam} constellations are used, \ie the set of available rates is defined as $\mcR=\big\{ \Rc M~|~M\in\set{4,6},\Rc\in (0,1] \big\}$. Here, $\Rc$ may be understood as a the rate of a binary encoder whose output is interleaved and then used by the modulator; thus we assume a \gls{bicm} is implemented \cite{Szczecinski_Alvarado_book}.

\begin{figure}[tb]
\psfrag{SNR}[cc][cc][\siz]{$\SNRav$~[dB]}
\psfrag{throughput}[cc][cc][\siz]{Throughput $\bpcu$}
\psfrag{ Erg}[lc][lc][\siz]{$\ov{C}$}
\psfrag{ AM}[lc][lc][\siz]{$\eta^{\amc}$}
\psfrag{ HA}[lc][lc][\siz]{$\eta^{\IR}_{4}$}
\psfrag{ LH2}[lc][lc][\siz]{$\eta^{\Lc}_{2}$}
\psfrag{ LH4}[lc][lc][\siz]{$\eta^{\Lc}_{4}$}
\psfrag{ LH2O}[lc][lc][\siz]{$\eta^{\Lc}_{2}(\hat{\Delta}_2)$}
\psfrag{ LH4O}[lc][lc][\siz]{$\eta^{\Lc}_{4}(\hat{\Delta}_4)$}
\begin{center}
\scalebox{\sizf}{\includegraphics[width=\linewidth]{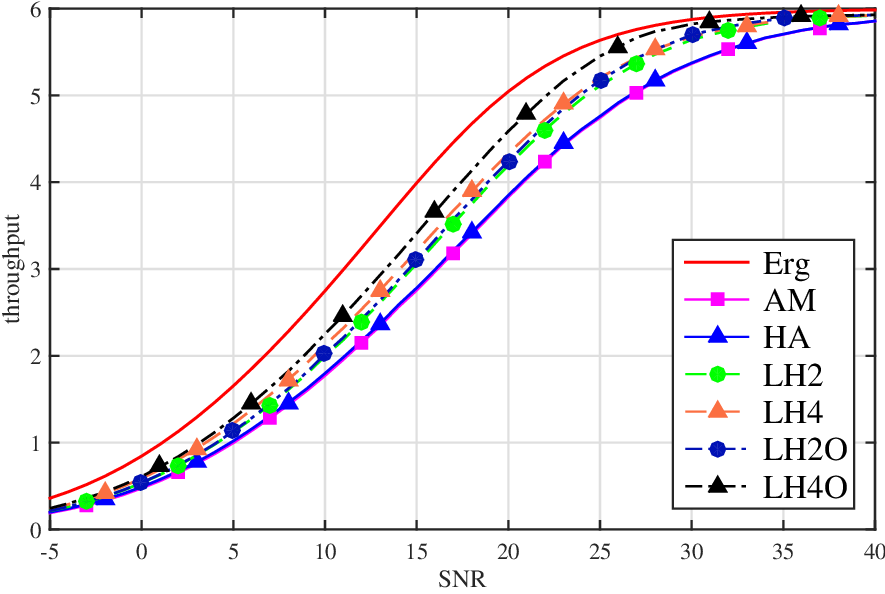}}\\
{\footnotesize a)}\\
\scalebox{\sizf}{\includegraphics[width=\linewidth]{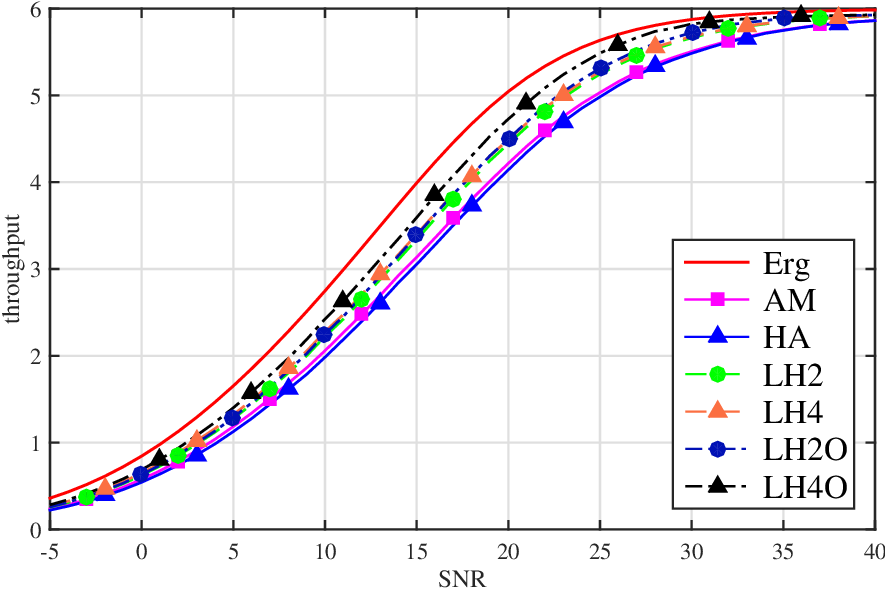}}\\
{\footnotesize b)}
\end{center}
\caption{$16$-\gls{qam} \& $64$-\gls{qam} transmission over Rayleigh fading channels; threshold decoding model. Throughput of the proposed \gls{lharq}, $\eta^{\Lc}_{\kmax}(\Delta)$, is compared to  the conventional \gls{irharq}, $\eta^{\IR}_{\kmax}$, when a)~$\fD \tau=0.1$, \ie $\corSNR\approx0.8$, and b)~$\fD \tau=0.05$, \ie $\corSNR\approx0.95$.}\label{Fig:Rayleigh.lharq0.95}
\end{figure}
We can see clearly that allowing modulation order adaptation does not change the conclusions drawn in \exref{Ex:QAM.Rayleigh}, \ie adding the conventional \gls{irharq} on top of \gls{amc} does not provide gains from throughput of view. However, the gains provided by \gls{lharq} over the conventional \gls{irharq} are notable already for $\kmax=2$. For instance, for a target throughput value of $\eta\in(2.5, 5) \bpcuval$, the \gls{snr} gap between the conventional \gls{irharq} and the ergodic capacity is reduced by approximately $2\dBval$. 
On the other hand, increasing the number of transmissions to $\kmax=4$ does not increase the gains. This is because the rate-adaptation policy of \gls{amc} is set irrespectively of the fact that \gls{harq} takes care of undecoded packets; it is thus defined conservatively to avoid ``too many'' decoding errors.

To take advantage of \gls{harq}, and improve the throughput, \gls{amc} must use a more aggressive rate adaptation; this can be done via fine-tuning defined in \eqref{delta.gamma}. The throughput of such aggressive \gls{lharq}, $\eta^{\Lc}_{\kmax}(\hat{\Delta})$ is also shown in \figref{Fig:Rayleigh.lharq0.95}, where we see that by allowing for more rounds, \eg $\kmax=4$, \gls{lharq} yields results even closer to the ergodic capacity, gaining another $1$--$2\dBval$. This is achieved by using rates that are higher than those indicated by \gls{amc} as shown in~\figref{Fig:Rayleigh.policy.lharq0.95}; we can see clearly that the optimal $\hat{\Delta}$ tends to increase when $\kmax$ increases. The transmission rate becomes more aggressive with larger $\kmax$ because more errors can be tolerated initially and the decoding is postponed to the later rounds. 

We observed that the optimal value $\hat{\Delta}$ increases slowly in low-to-medium $\SNRav$  but tends to increase faster for high \gls{snr}. For instance, in the case of $\kmax=4$ and $\fD \tau=0.05$, we obtained $\hat{\Delta}\in(1.2, 3.4)$ for $\SNRav\in(-5,20)\dBval$ while for $\SNRav=36\dBval$, we had $\hat{\Delta}\approx 50$. However, using $\Delta\approx 4$ even for high \gls{snr} does not cause a notable throughput penalty compared to  $\hat{\Delta}$. 

\end{example}

\begin{figure}[tb]
\psfrag{SNR}[cc][cc][\siz]{$\SNRav$~[dB]}
\psfrag{throughput}[cc][cc][\siz]{Throughput $\bpcu$}
\psfrag{ Erg}[lc][lc][\siz]{$\ov{C}$}
\psfrag{ AM}[lc][lc][\siz]{$\eta^{\amc}$}
\psfrag{ HA}[lc][lc][\siz]{$\eta^{\IR}_{4}$}
\psfrag{ LH2}[lc][lc][\siz]{$\eta^{\Lc}_{2}$}
\psfrag{ LH4}[lc][lc][\siz]{$\eta^{\Lc}_{4}$}
\psfrag{ LH2o}[lc][lc][\siz]{$\eta^{\Lc}_{2}(\hat{\Delta}_2)$}
\psfrag{ LH4o}[lc][lc][\siz]{$\eta^{\Lc}_{4}(\hat{\Delta}_4)$}
\psfrag{ VLH2}[lc][lc][\siz]{$\eta^{\VL}_{2}$}
\psfrag{ VLH4}[lc][lc][\siz]{$\eta^{\VL}_{4}$}
\psfrag{ VLH2o}[lc][lc][\siz]{$\eta^{\VL}_{2}(\hat{\Delta}_2)$}
\psfrag{ VLH4o}[lc][lc][\siz]{$\eta^{\VL}_{4}(\hat{\Delta}_4)$}
\psfrag{ XPH4o}[lc][lc][\siz]{$\eta^{\xp}_{4}(\hat{\Delta}_4)$}
\begin{center}
\scalebox{\sizf}{\includegraphics[width=\linewidth]{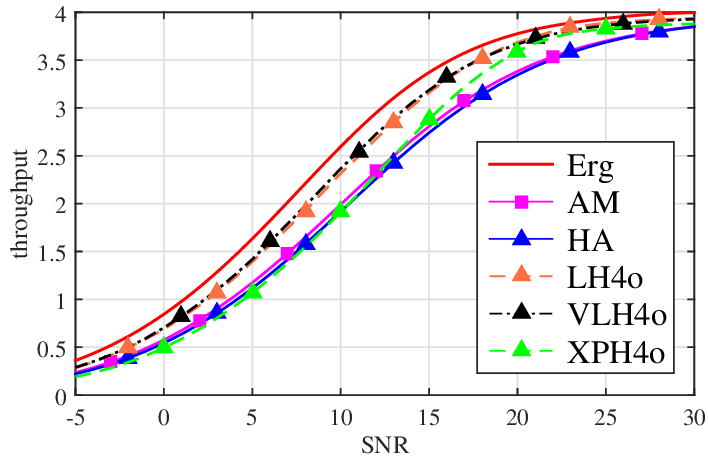}}\\
\end{center}
\caption{$16$-\gls{qam} transmission over Rayleigh fading channels; threshold decoding model. Throughput of the proposed \gls{lharq}, $\eta^{\Lc}_{\kmax}(\Delta)$, is compared to \gls{vlharq}, $\eta^{\VL}_{\kmax}(\Delta)$, to \gls{xp}, $\eta^{\xp}_{\kmax}(\Delta)$, and to the conventional \gls{irharq}, $\eta^{\IR}_{\kmax}$; $\fD \tau=0.05$, \ie $\corSNR\approx0.95$.}\label{Fig:vlharq.lharq0.95}
\end{figure}

\begin{example}[\gls{lharq} and previous works]\label{Ex:VL.vs.L}
The threshold decoding assumption used in the \exref{ex:lharq.rate.policy} provides us with the framework suitable for comparison with previous works which were also based on the similar assumption. Since, using various constellations does not alter the main conclusions we consider only $16$-\gls{qam} modulation \ie $\mcR=\big\{ 4\Rc~|~\Rc\in (0,1] \big\}$.

We start by comparing \gls{lharq}  and \gls{vlharq} introduced in \secref{Sec:VLHARQ}. As we said in \secref{Sec:VLHARQ} we have to devise a strategy to adapt the lengths $\ell_k, k=2,\ld, \kmax$ in each \gls{harq} round. Because we want the adaptation to be independent of the distribution of the \gls{csi}, we continue in the spirit of equation \eqref{eq:optimal/policy.K.1}. Namely, 
we propose to determine $R_1$ as in \gls{amc}, \ie exactly following \eqref{eq:optimal/policy.K.1}, and find $\ell_{2},\ld,\ell_{\kmax}$ by maximizing the per-round (instantaneous) throughput, which boils down to solving the following one-dimensional optimization problem:
\begin{align}\label{eq:optimal-length}
&\ell_{k}(\SNR_k)=\argmax_{\ell\in(0,1)}~\frac{\R_{1}(\SNR_1)}{\ell} \Big(1- \PER_k(\SNR_k, \ell;\cd)\Big), 
\end{align}
where $\PER_k(\SNR_k,\ell;\cd)$ is defined by \eqref{PER.function.MD} and, using the threshold decoding model, it is given by
\begin{align}
\PER\Big(\SNR_k, \ell;\set{\tilde{\SNR}_t}_{t=1}^{k-1},\R_{1},&\set{\ell_{t}}_{t=2}^{k-1}\Big)
\nonumber\\
\quad =\Pr\Big\{ \sum_{l=1}^{k-1} \ell_{l}(\SNR_l) I(\tilde{\SNR}_l)&+\ell I(\tilde{\SNRrv}_k)<\R_{1}(\SNR_1) \nonumber\\
&\big| \SNRrv_k=\SNR_k \Big\}
\end{align}
and can be find using Monte-Carlo simulations for any given $\R_1$ and $\SNR_k$. Despite the simplifying threshold decoding assumption, this is still a tedious exercise.

For completeness, we show also the results where the \gls{csi} is not available at the transmitter which should highlight the value of combining \gls{harq} and \gls{amc}. We use here the \gls{xp} derived in \cite[eq. (37)]{Jabi17a}.  

The results are shown in \figref{Fig:vlharq.lharq0.95}, where we can see that \gls{lharq} and \gls{vlharq} are practically the same for $\kmax=4$. It is also clear that the \gls{xp} without \gls{csi} is equivalent to the conventional \gls{irharq} at low-medium $\SNRav$. However, since in very high \gls{snr} regime the highest available rate is practically always used, \gls{xp} results (without \gls{csi}) are  the same as those of \gls{lharq} (with \gls{csi}).

We emphasize that, while we made the operation conditions as similar as possible, the results of \gls{vlharq} and \gls{xp} should be seen as  a ``ballpark'' figure rather than the rigorous comparison. As for \gls{vlharq}, this is because we do not consider any contraints on the lengths, \ie we ignore the issue of sharing the constant-length frame between many packets (of different lengths). As for \gls{xp}, the comparison is not fair neither because, in the absence of the instantaneous \gls{csi}, we adapt the length using the distribution of the \gls{csi}.
\end{example}

The  results in the previous example indicate that \gls{lharq} presents a notable gain compared to conventional \gls{irharq} assuming that {\it{i)}} the sets of transmission, $\mcR$, and backtrack, $\mcR_\tnr{b}$, rates are continuous, {\it{ii)}} the error event $\Err_{k}$ is fully described by the accumulated \gls{mi}, \cf \eqref{errk.mi}, which also allows us to determine the optimal backtrack rates $\Rs{k}$ in closed-form, see \eqref{rho.k.MI}. The purpose of the next example is to compare \gls{lharq} with \gls{irharq} when these idealized assumptions are abandoned and practical encoders/decoders are used.

We immediately say that we cannot follow the idea of comparing \gls{lharq} with  \gls{vlharq} or \gls{xp}. While we could do it in \exref{Ex:VL.vs.L}, this was done only thanks to the assumption of the idealized threshold decoding. The similar comparison is simply out of reach with current computational resources and time-constraints when we use the practical encoders and decoders.

\begin{figure}[tb]
\psfrag{SNR}[cc][cc][\siz]{$\SNRav$~[dB]}
\psfrag{throughput}[cc][cc][\siz]{Throughput $\bpcu$}
\psfrag{ Erg}[lc][lc][\siz]{$\ov{C}$}
\psfrag{ AM}[lc][lc][\siz]{$\eta^{\amc}$}
\psfrag{ HA}[lc][lc][\siz]{$\eta^{\IR}_{4}$}
\psfrag{ LH2}[lc][lc][\siz]{$\eta^{\Lc}_{2}$}
\psfrag{ LH4}[lc][lc][\siz]{$\eta^{\Lc}_{4}$}
\psfrag{ LH2Ox}[lc][lc][\siz]{$\eta^{\Lc}_{2}(\hat{\Delta})$}
\psfrag{ LH4O}[lc][lc][\siz]{$\eta^{\Lc}_{4}(\hat{\Delta})$}
\begin{center}
\scalebox{\sizf}{\includegraphics[width=\linewidth]{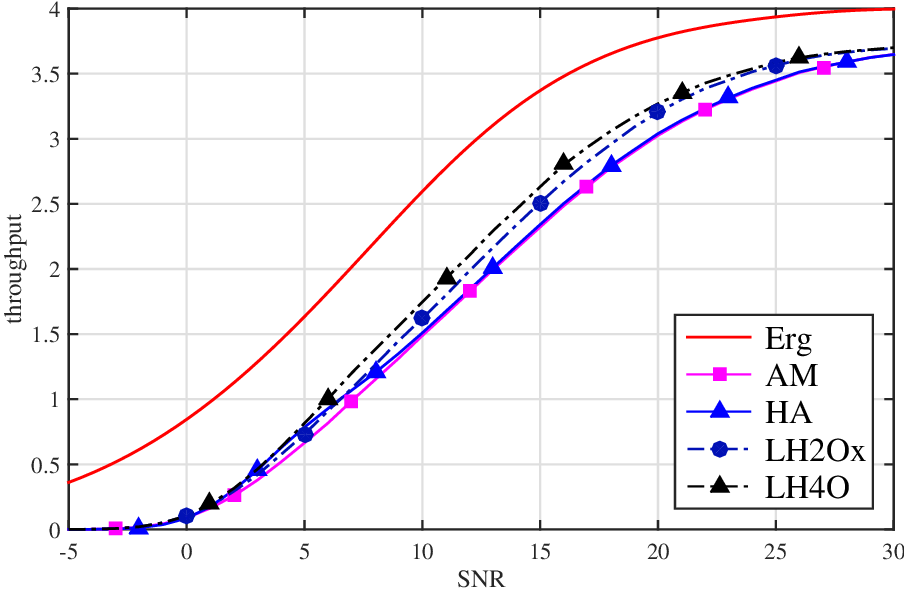}}\\
{\footnotesize a)}\\
\scalebox{\sizf}{\includegraphics[width=\linewidth]{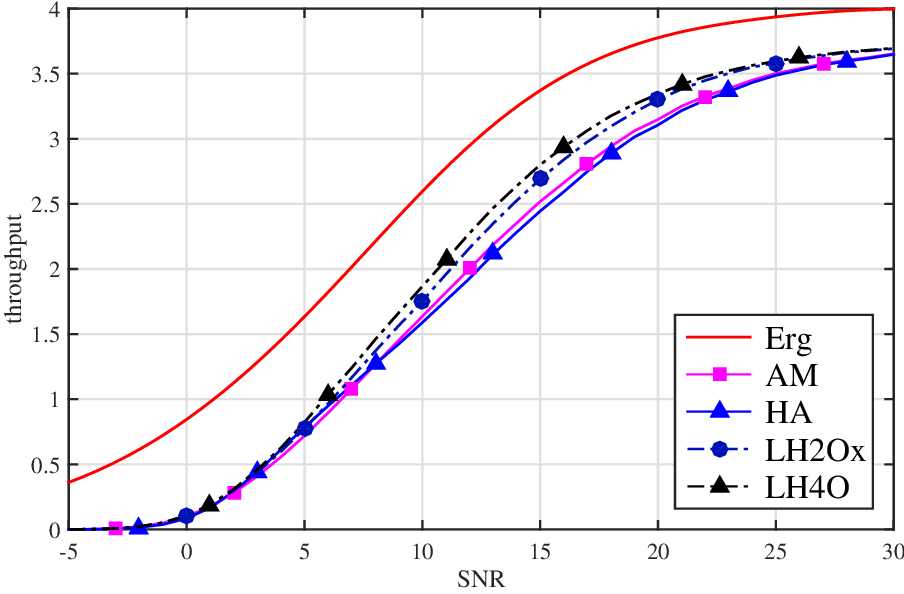}}\\
{\footnotesize b)}
\end{center}
\caption{Turbo-coded $16$-\gls{qam} transmission over Rayleigh fading channels: throughput of the proposed \gls{lharq}, $\eta^{\Lc}_{\kmax}(\Delta)$, compared to  the conventional \gls{irharq}, $\eta^{\IR}_{\kmax}$, when a)~$\fD \tau=0.1$, \ie $\corSNR\approx0.8$, and b)~$\fD \tau=0.05$, \ie $\corSNR\approx0.95$.}\label{Fig:Rayleigh.TC.lharq0.95}
\end{figure}

\begin{example}[Practical implementation using turbo codes]\label{Ex:TC}
We consider encoding of information packets using the $\frac{1}{3}$-rate \gls{3gpp} turbo-code followed by a \gls{3gpp} rate matching which allows us to obtain rates from \mbox{$\mcR=\set{1.5, 2.25, 3, 3.75}$}. The encoded bits are then mapped to symbols using a Gray mapping  \cite[Sec.~2.5.2]{Szczecinski_Alvarado_book}. The block size is $\Ns=1024$. In this example again, we use only $16$-\gls{qam} for $\mcX$ constellation (\ie \gls{amc} is used with fixed modulation, see footnote \ref{footnote:AMC}). At the receiver the \glspl{llr} are calculated \cite[Sec.~3.3]{Szczecinski_Alvarado_book} and fed to the 6-iterations binary decoder based on the \gls{bcjr} algorithm  implemented in the log domain using the library~\cite{Doray15}. The example of the \gls{per} curves obtained via simulations are shown in \figref{Fig:WEP_rate}. 


\begin{figure}[tb]
\psfrag{SNR}[cc][cc][\siz]{$\SNR$~[dB]}
\psfrag{WEP}[cc][cc][\siz]{$\Pr\{\Err|\SNRrv=\SNR,\R\}$}
\psfrag{R=1.5XX}[lc][lc][\siz]{$\R=1.5$}
\psfrag{R=2.25}[lc][lc][\siz]{$\R=2.25$}
\psfrag{R=3}[lc][lc][\siz]{$\R=3$}
\psfrag{R=3.75}[lc][lc][\siz]{$\R=3.75$}
\begin{center}
\scalebox{\sizf}{\includegraphics[width=\linewidth]{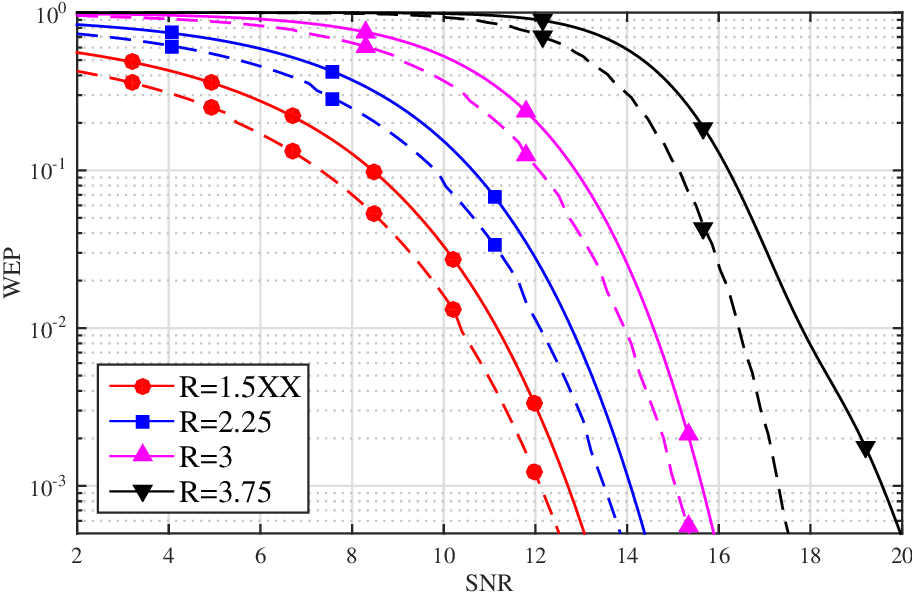}}
\end{center}
\caption{$\PER(\SNR;\R)=\PR{\Err|\SNRrv=\SNR,\R}$ is shown for different values of $\R$ and fading defined by \eqref{eq:cond.probablity} with $\delta=0.95$, \ie $f_\tr{D}\tau=0.05$ and $\SNRav=15$dB. The solid lines correspond to a turbo-code transmission with the $16$-\gls{qam} we use in \exref{Ex:TC}, while dashed ones -- to the idealized, threshold decoding we use in \exref{ex:lharq.rate.policy}.}\label{Fig:WEP_rate}
\end{figure}

After each \gls{nack} message, $\mfm'_{[k]}$ is a punctured version of $\mfm_k$, \ie $\mfm'_{[k]}=\Phi_k^\tnr{b}[\mfm_k]$ which means that $\rho_{k}\le \R_{k}$ by construction. For each available transmission rate $\R$, we use uniformly distributed backtrack rates. Namely, we consider $\mcR_\tnr{b}(R)=\set{k\R/F, k\in\set{0,1,\ld,F}}$ where $\log_2(F)$ is the number of the additional feedback bits needed to send the index of the chosen $\Rs{k}$ to the transmitter. In the numerical example, we choose $\log_2(F)=4$. 
 
\figref{Fig:WEP_backtrack_rate} shows $\Pr\{\Err_{k}\wedge \Errb{k}|\tilde{\SNRrv}_k=\tilde{\SNR}_k,\R_k,\Rs{k}\}$ as a function of the \gls{snr} experienced in the $k$-th round, $\tilde{\SNR}_{k}$ for different values of $\Rs{k}$ when $\R_k=3$. We recall that the event \mbox{$\set{\Err_{k}\wedge \Errb{k}}$} means that both the ``direct'' and the ``backtrack'' decoding of the packet $\mfm_k$ failed. It is clear that as the number of known bits $\mfm'_{[k]}$ increases, \ie when $\Rs{k}$ increases, the probability of  a failure in  the backtrack decoding decreases. We note that when $\Rs{k}=0$ we have $\Err_{k}=\Errb{k}$. 
 
\begin{figure}[tb]
\psfrag{SNR}[cc][cc][\siz]{$\tilde{\SNR}_k$~[dB]}
\psfrag{WEP}[cc][cc][\siz]{$\PR{\Err_{k}\wedge \Errb{k}|\tilde{\SNRrv}_k=\tilde{\SNR}_k,\R_k,\Rs{k}}$}
\psfrag{R=0.0XX}[lc][lc][\siz]{$\Rs{k}=0$}
\psfrag{R=0.5}[lc][lc][\siz]{$\Rs{k}=0.5$}
\psfrag{R=1.3}[lc][lc][\siz]{$\Rs{k}=1.3$}
\psfrag{R=1.8}[lc][lc][\siz]{$\Rs{k}=1.8$}
\psfrag{R=2.6}[lc][lc][\siz]{$\Rs{k}=2.6$}
\begin{center}
\scalebox{\sizf}{\includegraphics[width=\linewidth]{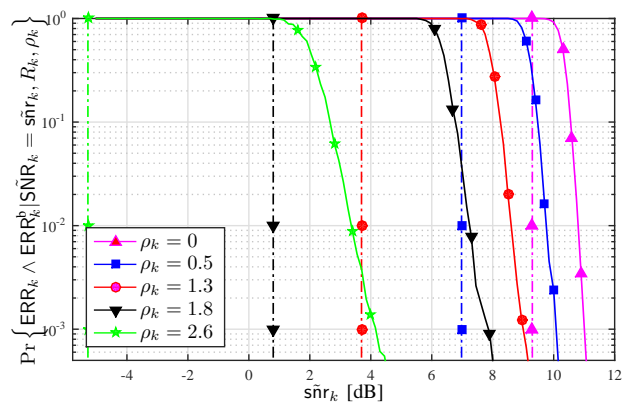}}
\end{center}
\caption{$\Pr\{\Err_{k}\wedge \Errb{k}|\tilde{\SNRrv}_k=\tilde{\SNR}_k,\R_k,\Rs{k}\}$, shown for $\R_k=3$ and different values of $\Rs{k}$, allows us to calculate the backtrack \gls{per} function \eqref{PER.back.2}.  The solid lines correspond to a turbo-code transmission wih $16$-\gls{qam} as in \exref{Ex:TC}, while dashed ones -- to the idealized, threshold decoding we use in \exref{ex:lharq.rate.policy}.}\label{Fig:WEP_backtrack_rate}
\end{figure}

The mixing rate policy is defined via \eqref{rho.1.heuristic} with $\epsilon_\tnr{b} = 0.1$, which follows \cite[Sec. V.B]{Jabi17b} and the results of \gls{lharq} are compared to \gls{irharq} in \figref{Fig:Rayleigh.TC.lharq0.95}a and \figref{Fig:Rayleigh.TC.lharq0.95}b  for $\fD \tau=0.1$ and $\fD \tau=0.05$, respectively. As already highlighted in \exref{Ex:QAM.Rayleigh}, \gls{irharq} is counterproductive in the region of high values of \gls{snr}, especially when the estimated $\SNR_k$ tends to be reliable (large Doppler, $\fD \tau=0.05$); here, this conclusion is confirmed in the practical setup as well. Furthermore, the gains of \gls{lharq} with respect to \gls{irharq}, theoretically predicted, still materialize in the considered practical scenarios. For instance, when $\fD \tau=0.1$, the \gls{snr} gain of \gls{lharq} over the conventional \gls{irharq} is around $1.5\dBval$ for a throughput equal to $\eta=3 \bpcuval$ and $\kmax=2$. This gain can be increased by using $\kmax=4$, but the improvement is less important than in the case of the idealized threshold decoding we have shown in \exref{ex:lharq.rate.policy}.

This can be understood observing \figref{Fig:WEP_backtrack_rate} where we see that for the same value of the \gls{snr} $\tilde{\SNR}_k$, to make the idealized threshold decoding succeed, the required value of the mixing rate $\Rs{k}$ is smaller than in the case of the practical codes. This is measured by the \gls{snr}-shift of the curve $\PR{\Err_{k}\wedge \Errb{k}|\tilde{\SNRrv}_k=\tilde{\SNR}_k,\R_k,\Rs{k}}$ with respect to the curve $\PR{\Err_{k}\wedge \Errb{k}|\tilde{\SNRrv}_k=\tilde{\SNR}_k,\R_k, 0}$; for example, the idealized threshold decoding curve is shifted to the left by $\sim2.5\dBval$ for $\Rs{k}=0.5$, while using the turbo-code, this shift is only equal to $\sim 1\dBval$. Smaller shift means that higher values of $\Rs{k}$ must be used to provide the similar guarantees for a successful backtrack decoding, and this translates also into a smaller throughput.

This effect highlights the importance of the suitable coding design which takes into account the reality of the backtrack decoding as it was studied in \cite[Sec.~IV.B]{Jabi17b}.

We also observed that the aggressivity factor $\hat{\Delta}$ tends to be much smaller in the case of practical decoder/encoder. For instance,  $\hat{\Delta}\le 2$ for all values of \gls{snr} when $\kmax=4$ and $\fD \tau=0.05$.
\end{example}

\section{Conclusions}\label{Sec:Conclusions}

This work is motivated by the fact that combining the conventional \gls{harq} with \gls{amc} transmissions over \gls{iid} block fading channels is detrimental to the throughput when the transmission resources are fixed. As a remedy to this problem, we proposed and analyzed a coding strategy designed to increase the throughput of \gls{harq} allowing it to exploit the \gls{csi} observed at the receiver. The proposed coding strategy is simple to implement because the coding is separated into two logical steps: the channel coding as done in the conventional \gls{amc} and the packet mixing done at the bit-level which may be associated with \gls{harq}. 

The optimization problem, formulated to find the throughput-maximizing rate adaptation functions, turns out to be infeasibly complex to solve. The heuristic solutions are then proposed and analyzed on examples. Using the information-theoretic approach to coding, and Rayleigh block-fading channels, the throughput obtained thanks to the proposed \gls{lharq}, shows a gain of $2\dBval$ to $4\dBval$ when compared to the conventional \gls{amc}. In a similar setup, but using the practical, \gls{3gpp} turbo-code, the  proposed solution shows notable gains of $2\dBval$ to $2.5\dBval$.  Since these gains are obtained using additional bits in the feedback channel, the tradeoff between the additional signaling and the throughput improvement should be further assessed using system-level considerations.

We conjecture also that the gap between the practical codes and the theoretical limits is due to the particular coding strategy adopted here. The optimization of the coding scheme which would take into account the structure of the decoding scheme, would improve the results as already shown in \cite{Jabi17b}.

Since the proposed adaptation strategy maintains constant resources allocated to each user, we argue that it may be particularly suitable to combine the analysis with the multi-user resource management procedures, a problem which is left for further investigation. Moreover, it would be interesting to compare \gls{vlharq} and \gls{lharq} from a system point of view, taking into account additional practical constraints such as feedback costs and channel resource granularity.



\balance

\bibliography{IEEEabrv,references_all}

\begin{thebibliography}{10}
\providecommand{\url}[1]{#1}
\csname url@samestyle\endcsname
\providecommand{\newblock}{\relax}
\providecommand{\bibinfo}[2]{#2}
\providecommand{\BIBentrySTDinterwordspacing}{\spaceskip=0pt\relax}
\providecommand{\BIBentryALTinterwordstretchfactor}{4}
\providecommand{\BIBentryALTinterwordspacing}{\spaceskip=\fontdimen2\font plus
\BIBentryALTinterwordstretchfactor\fontdimen3\font minus
  \fontdimen4\font\relax}
\providecommand{\BIBforeignlanguage}[2]{{%
\expandafter\ifx\csname l@#1\endcsname\relax
\typeout{** WARNING: IEEEtran.bst: No hyphenation pattern has been}%
\typeout{** loaded for the language `#1'. Using the pattern for}%
\typeout{** the default language instead.}%
\else
\language=\csname l@#1\endcsname
\fi
#2}}
\providecommand{\BIBdecl}{\relax}
\BIBdecl

\bibitem{Sassioui17}
R.~Sassioui, M.~Jabi, L.~Szczecinski, L.~B. Le, M.~Benjillali, and
  B.~Pelletier, ``{HARQ} and {AMC}: Friends or foes?'' \emph{{IEEE} Trans.
  Commun.}, vol.~65, no.~2, pp. 635--650, Feb. 2017.

\bibitem{Popovski14}
P.~Popovski, ``Delayed channel state information: Incremental redundancy with
  backtrack retransmission,'' in \emph{{IEEE} Inter. Conf. Comm. (ICC)}, June
  2014, pp. 2045--2051.

\bibitem{Jabi17b}
M.~Jabi, E.~Pierre-Doray, L.~Szczecinski, and M.~Benjillali, ``How to boost the
  throughput of {HARQ} with off-the-shelf codes,'' \emph{{IEEE} Trans.
  Commun.}, vol.~65, no.~6, pp. 2319--2331, June 2017.

\bibitem{Liu04}
Q.~Liu, S.~Zhou, and G.~B. Giannakis, ``Cross-layer combining of adaptive
  modulation and coding with truncated {ARQ} over wireless links,''
  \emph{{IEEE} Trans. Wireless Commun.}, vol.~3, no.~5, pp. 1746--1755, Sep.
  2004.

\bibitem{Zheng05}
H.~Zheng and H.~Viswanathan, ``Optimizing the {ARQ} performance in downlink
  packet data systems with scheduling,'' \emph{{IEEE} Trans. Wireless Commun.},
  vol.~4, no.~2, pp. 495--506, Mar. 2005.

\bibitem{Kim08b}
D.~Kim, B.~C. Jung, H.~Lee, D.~K. Sung, and H.~Yoon, ``Optimal modulation and
  coding scheme selection in cellular networks with hybrid-{ARQ} error
  control,'' \emph{{IEEE} Trans. Wireless Commun.}, vol.~7, no.~2, pp.
  5195--5201, Dec. 2008.

\bibitem{Ramis11}
J.~Ramis and G.~Femenias, ``Cross-layer design of adaptive multirate wireless
  networks using truncated {HARQ},'' \emph{{IEEE} Trans. Veh. Technol.},
  vol.~60, no.~3, pp. 944--954, Mar. 2011.

\bibitem{Zhang13}
P.~Zhang, Y.~Miao, and Y.~Zhao, ``Cross-layer design of {AMC} and truncated
  {HARQ} using dynamic switching thresholds,'' in \emph{{IEEE} Wireless
  Communications and Networking Conference ({WCNC}'13), 7-10 April, Shanghai,
  China}, Apr. 2013, pp. 906--911.

\bibitem{Kang10}
C.~G. Kang, S.~H. Park, and J.~W. Kim, ``Design of adaptive modulation and
  coding scheme for truncated hybrid arq,'' \emph{Wireless Personal
  Communications}, vol.~53, no.~2, pp. 269--280, Apr. 2010.

\bibitem{Le06}
L.~Le, E.~Hossain, and A.~Alfa, ``Service differentiation in multirate wireless
  networks with weighted round-robin scheduling and {ARQ}-based error
  control,'' \emph{{IEEE} Trans. Commun.}, vol.~54, no.~2, pp. 208--215, Feb.
  2006.

\bibitem{Wang07}
X.~Wang, Q.~Liu, and G.~B. Giannakis, ``Analyzing and optimizing adaptive
  modulation coding jointly with {ARQ} for {QoS}-guaranteed traffic,''
  \emph{{IEEE} Trans. Veh. Technol.}, vol.~56, no.~2, pp. 710--720, Mar. 2007.

\bibitem{Le07}
L.~Le, E.~Hossain, and M.~Zorzi, ``Queueing analysis for {GBN} and {SR} {ARQ}
  protocols under dynamic radio link adaptation with non-zero feedback delay,''
  \emph{{IEEE} Trans. Wireless Commun.}, vol.~54, no.~9, pp. 3418--3428, Feb.
  2007.

\bibitem{Femenias09}
G.~Femenias, J.~Ramis, and L.~Carrasco, ``Using two-dimensional {M}arkov models
  and the effective-capacity approach for cross-layer design in
  {AMC}/{ARQ}-based wireless networks,'' \emph{{IEEE} Trans. Veh. Technol.},
  vol.~58, no.~8, pp. 4193--4203, Oct. 2009.

\bibitem{Poggioni10}
M.~Poggioni, L.~Rugini, and P.~Banelli, ``{QoS} analysis of a scheduling policy
  for heterogeneous users employing {AMC} jointly with {ARQ},'' \emph{{IEEE}
  Trans. Commun.}, vol.~58, no.~9, pp. 2639--2652, September 2010.

\bibitem{Carrasco2013}
L.~Carrasco, G.~Femenias, and J.~Ramis, ``Channel-aware {MAC} performance of
  {AMC}-{ARQ}-based wireless systems,'' \emph{{EURASIP} Journal on Wireless
  Communications and Networking}, vol. 2013, no.~1, p. 213, Aug 2013.

\bibitem{Harsini11}
J.~Harsini, F.~Lahouti, M.~Levorato, and M.~Zorzi, ``Analysis of
  non-cooperative and cooperative type {II} hybrid {ARQ} protocols with {AMC}
  over correlated fading channels,'' \emph{{IEEE} Trans. Wireless Commun.},
  vol.~10, no.~3, pp. 877 --889, Mar. 2011.

\bibitem{Caire01}
G.~Caire and D.~Tuninetti, ``The throughput of hybrid-{ARQ} protocols for the
  {G}aussian collision channel,'' \emph{{IEEE} Trans. Inf. Theory}, vol.~47,
  no.~5, pp. 1971--1988, Jul. 2001.

\bibitem{Sassioui16}
R.~Sassioui, L.~Szczecinski, L.~B. Le, and M.~Benjillali, ``{AMC} and {HARQ}:
  Effective capacity analysis,'' in \emph{{IEEE} Wireless Communications and
  Networking Conference ({WCNC}'16), 3-6 April, Doha, Qatar}, 2016.

\bibitem{Larsson14}
P.~Larsson, L.~K. Rasmussen, and M.~Skoglund, ``Throughput analysis of {ARQ}
  schemes in {G}aussian block fading channels,'' \emph{{IEEE} Trans. Commun.},
  vol.~62, no.~7, pp. 2569--2588, Jul. 2014.

\bibitem{Jabi15b}
M.~Jabi, A.~{El Hamss}, L.~Szczecinski, and P.~Piantanida, ``Multi-packet
  hybrid {ARQ}: Closing gap to the ergodic capacity,'' \emph{{IEEE} Trans.
  Commun.}, vol.~63, no.~12, pp. 5191--5205, Dec. 2015.

\bibitem{Jabi17a}
M.~Jabi, A.~Benyouss, M.~L. Treust, E.~Pierre-Doray, and L.~Szczecinski,
  ``Adaptive cross-packet {HARQ},'' \emph{{IEEE} Trans. Commun.}, vol.~65,
  no.~5, pp. 2022--2035, May 2017.

\bibitem{Cheng03}
J.-F. Cheng, Y.-P. Wang, and S.~Parkvall, ``Adaptive incremental redundancy,''
  in \emph{{IEEE} Veh. Tech. Conf. (VTC Fall)}, Orlando, Florida, USA, Oct.
  2003, pp. 737--741.

\bibitem{Uhlemann03}
E.~Uhlemann, L.~K. Rasmussen, A.~Grant, and P.-A. Wiberg, ``Optimal
  incremental-redundancy strategy for type-{II} hybrid {ARQ},'' in \emph{{IEEE}
  Intern. Symp. Inf. Theory {(ISIT)}}, 2003, p. 448.

\bibitem{Visotsky05}
E.~Visotsky, Y.~Sun, V.~Tripathi, M.~Honig, and R.~Peterson,
  ``Reliability-based incremental redundancy with convolutional codes,''
  \emph{{IEEE} Trans. Commun.}, vol.~53, no.~6, pp. 987--997, Jun. 2005.

\bibitem{Pfletschinger10}
S.~Pfletschinger and M.~Navarro, ``Adaptive {HARQ} for imperfect channel
  knowledge,'' in \emph{2010 International {ITG} Conference on Source and
  Channel Coding ({SCC})}, Jan. 2010, pp. 1--6.

\bibitem{Szczecinski13}
L.~Szczecinski, S.~R. Khosravirad, P.~Duhamel, and M.~Rahman, ``Rate allocation
  and adaptation for incremental redundancy truncated {HARQ},'' \emph{{IEEE}
  Trans. Commun.}, vol.~61, no.~6, pp. 2580--2590, June 2013.

\bibitem{Jabi16}
M.~Jabi, M.~Benjillali, L.~Szczecinski, and F.~Labeau, ``Energy efficiency of
  adaptive {HARQ},'' \emph{{IEEE} Trans. Commun.}, vol.~64, no.~2, pp.
  818--831, Feb. 2016.

\bibitem{Verdu10}
S.~Verdu and S.~Shamai, ``Variable-rate channel capacity,'' \emph{{IEEE} Trans.
  Inf. Theory}, vol.~56, no.~6, pp. 2651--2667, Jun. 2010.

\bibitem{Dahlman14_book}
E.~Dahlman, S.~Parkval, and J.~Skold, \emph{4{G} {LTE}/{LTE}-advanced for
  mobile broadband}, 2nd~ed.\hskip 1em plus 0.5em minus 0.4em\relax Academic
  Press, 2014.

\bibitem{Trillingsgaard17}
K.~F. Trillingsgaard and P.~Popovski, ``Generalized {HARQ} protocols with
  delayed channel state information and average latency constraints,''
  \emph{{IEEE} Trans. Inf. Theory}, vol.~PP, no.~99, pp. 1--1, 2017.

\bibitem{3GPP_TS_36.213}
3GPP, ``{3GPP} {TS} 36.213 {V}13.1.1 release 13 (2016-03): {LTE}; {E}volved
  {U}niversal {T}errestrial {R}adio {A}ccess ({E-UTRA}); {P}hysical layer
  procedures.''

\bibitem{Alouini00}
M.-S. Alouini and A.~J. Goldsmith, ``Adaptive modulation over {Nakagami} fading
  channels,'' \emph{Kluwer Journal on Wireless Communications}, vol.~13, no.
  1-2, pp. 119--143, May 2000.

\bibitem{Levorato09}
M.~Levorato, L.~Badia, and M.~Zorzi, ``On the channel statistics in hybrid
  {ARQ} systems for correlated channels,'' in \emph{{IEEE} Information Theory
  Workshop ({ITW})}, Oct. 2009, pp. 90--94.

\bibitem{Hu13}
Y.~Hu and A.~Ribeiro, ``Optimal wireless communications with imperfect channel
  state information,'' \emph{{IEEE} Trans. Signal Process.}, vol.~61, no.~11,
  pp. 2751--2766, Jun. 2013.

\bibitem{Proakis08_Book}
J.~G. Proakis and M.~Salehi, \emph{Digital Communications}, 5th~ed.\hskip 1em
  plus 0.5em minus 0.4em\relax McGraw-Hill, 2008.

\bibitem{Wu10}
P.~Wu and N.~Jindal, ``Performance of hybrid-{ARQ} in block-fading channels: A
  fixed outage probability analysis,'' \emph{{IEEE} Trans. Commun.}, vol.~58,
  no.~4, pp. 1129--1141, Apr. 2010.

\bibitem{Szczecinski_Alvarado_book}
L.~Szczecinski and A.~Alvarado, \emph{Bit-Interlaved Coded Modulation :
  Fundamentals, Analysis and Design}.\hskip 1em plus 0.5em minus 0.4em\relax
  Wiley, 2015.

\bibitem{Ellis15}
J.~D. Ellis and M.~B. Pursley, ``Integration of adaptive modulation and channel
  coding with fountain coding for packet radio systems,'' \emph{{IEEE} Trans.
  Commun.}, vol.~63, no.~5, pp. 1510--1521, May 2015.

\bibitem{Bertsekas05_book}
D.~P. Bertsekas, \emph{Dynamic Programming and Optimal Control}, 3rd~ed.\hskip
  1em plus 0.5em minus 0.4em\relax Athena Scientific, 2005, vol.~2.

\bibitem{Doray15}
\BIBentryALTinterwordspacing
E.~Pierre-Doray and L.~Szczecinski. (2015) ``{FeCl} channel coding library''.
  [Online]. Available: \url{https://github.com/eti-p-doray/FeCl/wiki}
\BIBentrySTDinterwordspacing

\end{thebibliography}
\bibliographystyle{IEEEtran}


\end{document}